\documentclass[aip,jcp,floatfix,10pt,twocolumn,notitlepage]{revtex4-1}

\usepackage{graphicx}
\usepackage{amssymb,amsmath,amsbsy,amsfonts,bm}
\usepackage{multirow}
\usepackage{tikz}
\usepackage{hyperref}
\usepackage{gensymb}
\usepackage{threeparttable}
\usepackage{csquotes}
\usepackage{braket}
\usepackage[capitalize]{cleveref}
\usepackage{comment}
\usepackage[percent]{overpic}
\usepackage{color}

\newlength{\myl}
\newcommand{\bvec}[1]{{\mathbf{\string#1} }}
\newcommand{\upd}{\mathrm{d}}

\newcommand{\SUM}[2]{{\setlength{\myl}{\widthof{$\displaystyle\sum_{#1}^{#2}$}*\real{0.5}-\widthof{$\displaystyle\sum$}*\real{0.5}}\sum_{#1}^{#2}\;\hspace{-\the\myl}}}
\newcommand{\INT}[3]{\settowidth{\myl}{$\displaystyle\int_{#1}^{#2}$}{\int_{#1}^{#2}\;\;\;\hspace{-\the\myl}\dif #3}\,}

\begin{document}
\title{Biaxial nematic order in fundamental measure theory}

\author{Anouar El Moumane}
\affiliation{{Institut f\"ur Theoretische Physik II: Weiche Materie, Heinrich-Heine-Universit\"at D\"usseldorf, 40225 D\"usseldorf, Germany}}

\author{Michael te Vrugt}
\affiliation{DAMTP, Centre for Mathematical Sciences, University of Cambridge, Cambridge CB3 0WA, United Kingdom}

\author{Hartmut L\"owen}
\affiliation{{Institut f\"ur Theoretische Physik II: Weiche Materie, Heinrich-Heine-Universit\"at D\"usseldorf, 40225 D\"usseldorf, Germany}}

\author{Ren\'e Wittmann}
\email{rene.wittmann@hhu.de}
\affiliation{{Institut f\"ur Theoretische Physik II: Weiche Materie, Heinrich-Heine-Universit\"at D\"usseldorf, 40225 D\"usseldorf, Germany}}
\affiliation{Institut für Sicherheit und Qualität bei Fleisch, Max Rubner-Institut, 95326 Kulmbach, Germany}
\date{\today}

\begin{abstract}
Liquid crystals consisting of biaxial particles can exhibit a much richer phase behavior than their uniaxial counterparts. Usually, one has to rely on simulation results to understand the phase diagram of these systems, since very few analytical results exist. In this work, we apply fundamental measure theory, which allows us to derive free energy functionals for hard particles from first principles and with high accuracy, to systems of hard cylinders, cones and spherotriangles. We provide a general recipe for incorporating biaxial liquid crystal order parameters into fundamental measure theory and use this framework to obtain the phase boundaries for the emergence of orientational order in the considered systems. Our results provide insights into the phase behavior of biaxial nematic liquid crystals and, in particular, into methods for their analytical investigation.
\end{abstract}

\maketitle

    \section{Introduction}
The rich collective behavior that results from anisotropic interactions between particles has  fascinated liquid crystal researchers for decades~\cite{PhLiqCrys}.
 Even in spatially homogeneous systems, the particle's orientations can be uncorrelated (\textit{isotropic phase}) or correlated (\textit{ordered phases}).
A typical example for an orientationally ordered phase is the \textit{nematic phase}, in which the main symmetry axis of the particles is aligned.
Over the years, more and more phases with nontrivial ordering behavior have been identified~\cite{sebastian2022ferroelectric,Bruce2004}.
Prominent examples that have attracted interest in recent years are ferroelectric nematics (global polar order in which additional top-down symmetry is broken) \cite{chen2020first,lavrentovich2020ferroelectric} and biaxial nematics (two main axes of preferred orientation) \cite{freiser1970ordered,madsen2004thermotropic,LuckhurstS2015}.
Beyond these homogeneous ordered phases, many systems can also exhibit additional positional order, forming inhomogeneous liquid crystal phases such as columnars and smectics.

 Liquid crystalline order has also been observed in colloidal systems \cite{vroege1992phase,Smalyukh2018},
whose constituting particles can be synthesized nowadays with nearly any desired shape \cite{sacanna2013engineering,hueckel2021total}.
The entropic nature of the interactions governing these systems allows for a reliable theoretical modeling in terms of pure hard-core interactions~\cite{Mederos2014}.
Therefore, the ordering behavior of various hard-body systems
is well studied, in particular, for apolar uniaxial shapes such as rods or disks \cite{mcgrother1996re,bolhuis1997tracing,wensink2003isotropic,pfleiderer2007simple,marechal2011phase,odriozola2012revisiting}.
Moreover, while shape polarity is typically not sufficient to stabilize polar order in a homogeneous phase of purely hard particles, it can result in a rich behavior including polar domains at higher densities \cite{kubala2023splay} or even periodic gyroidal structures for particular pear-like shapes \cite{ellison2006entropy,schonhofer2017purely}.
Arguably, biaxial shapes give rise to an even more complex phase behavior due to their lower symmetry.
Even in two spatial dimensions, ordering phenomena related to two distinct particle axes reach from interlocked layers \cite{monderkamp2023network} to the emergence of nontrivial orientational symmetries \cite{martinez2021failure,martinez2022effect}.
In three spatial dimensions, the biaxial ordering of board-like particles (cuboids or spheroplatelets)
is well studied \cite{alben1973phase,taylor1991nematic,vanakaras2003theory,martinez2011biaxial,belli2011polydispersity,belli2012depletion,peroukidis2013phase,cuetos2017phase}.
These shapes are the common choice for investigating biaxiality due to their particular symmetry and their realization in colloidal experiments \cite{van2009experimental}.
However, quantitative theoretical results remain scarce.

The phase behavior of systems with complex particle shapes is typically investigated using computer simulations.
For gaining a deeper understanding of the phase diagram of complex liquid crystals and for reducing the computational effort, it is advantageous to have available an analytic theory which also represents the actual geometric interactions between the particles.
The best candidate for such a theoretical description is fundamental measure theory (FMT) \cite{Rosenfeld1989,roth2010fundamental,roth2012communication}, which allows us to obtain quantitatively accurate free energy functionals for hard particle systems from first principles.
Following its generalization from hard spheres to general convex hard particles \cite{hansen2009edFMT,hansen2010tensorial}, it has been successfully  refined \cite{wittmann2014,wittmann2015FMMT,wittmann2015THESIS,wittmann2016,marechal2017density} and applied to study the bulk phase behavior and various inhomogeneous
systems of (mostly) uniaxial rod-like particles \cite{marechal2013density,wittmann2014surface,wittmann2014,wittmann2015FMMT,wittmann2015,wittmann2016,marechal2017density,schonhofer2018density}.
 For hard spherocylinders in particular, FMT has been tested in detail and proven very reliable for a large range of problems \cite{wittmann2016}.
A prominent example for polar particles investigated by FMT are those of pear-like shape \cite{schonhofer2018density}.
Although triangular prisms have also been studied using FMT \cite{marechal2017density},
 they were implicitly assumed as uniaxial and only shapes were considered at which no biaxial order was found in simulations.
In turn, a particular version of FMT for parallel hard parallelepipeds was used to treat biaxiality under the assumption of restricted orientations \cite{martinez2011biaxial}.

\begin{figure}[t]
\centering
    \includegraphics[width=0.45\textwidth]{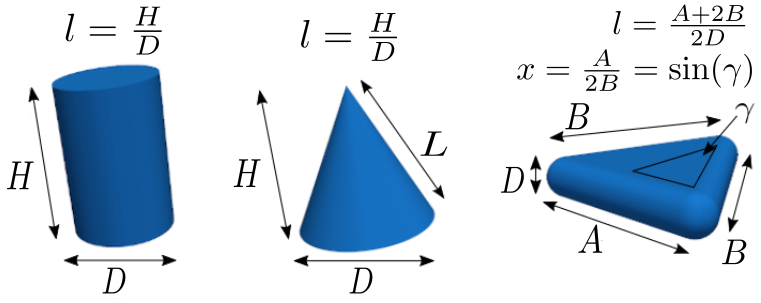}
    \caption{Parameterizations of the hard uniaxial and biaxial bodies considered in this work. Uniaxial cylinders (left) and cones (middle) are characterized by their (base) diameter $D$ and height $H$, which determine their common aspect ratio $l=H/D$.
    The mantle length $L$ of a cylinder ($L=H$)  differs from that of a cone ($L=\sqrt{H^2+D^2/4}$).
    Biaxial isosceles spherotriangles (right) of diameter $D$ are
    the parallel sets at distance $D/2$ of flat isosceles triangles with base length $A$ and side lengths $B$ (i.e., particles comprised of all points with distance equal or less than $D/2$ to a flat isosceles triangle).
        We define their aspect ratio $l=(A+2B)/(2D)$, their shape ratio $x=A/(2B)$ and their opening angle $2\gamma$, such that $x=\sin\gamma$.
    \label{fig_bodies}}
\end{figure}

The general strength of FMT is the possibility to gain analytic
insight into the phase transitions between ordered phases, because it directly connects anisotropic shape information to the particle orientation.
Analytic theories describing orientational ordering phenomena often rely on the use of \textit{orientational order parameters}, which measure the degree of orientational order in a system.
The most widely used one is the nematic order parameter measuring alignment of rod-like particles.
In systems of particles with complex shapes, different types of orientational order are possible, such that it is useful to employ a set of several order parameters \cite{rosso2007orientational,te2020relations} that measures orientational order with respect to different particle axes.
While it is well understood by now how the free energy for a homogeneous fluid of uniaxial particles can be expressed as a function of the average number density and the nematic order parameter~\cite{hansen2010tensorial,wittmann2014,wittmann2017twodim}, the free energy and, consequently, the phase behavior of biaxial particles as a function of appropriate orientational order parameters remains to be investigated.
In particular, it is not known how these order parameters can be incorporated into FMT.

In this work, we apply FMT to hard particles with uniaxial, polar and biaxial shapes, specifically to cylinders, cones and spherotriangles, as illustrated in Fig.~\ref{fig_bodies}, taking into account all orientational degrees of freedom.
We demonstrate how shape polarity affects the isotropic--nematic phase boundary by comparing a polar hard cone to an apolar hard cylinder
and provide detailed phase diagrams for the homogeneous phases of hard isosceles spherotriangles, also resolving the transitions between different uniaxial nematic phases and a biaxial phase.
In doing so, we also provide in Sec.~\ref{sec_theory} a generally applicable recipe for how to incorporate a certain set of relevant order parameters into FMT.
This general theory allows us to
overcome the need to use different assumptions to describe different phase transitions,
while still providing analytic insight.
These results are discussed in Sec.~\ref{sec_results}.

\section{Theory of orientational order \label{sec_theory}}

\subsection{Order parameters \label{sec_SUPF}}

\subsubsection{Uniaxial and biaxial particles}
The orientation of a hard particle in three spatial dimensions can in general be specified using a set of three angles $\theta \in [0,\pi]$, $\phi\in[0,2\pi]$ and $\psi\in[0,2\pi]$, known as the \textit{Euler angles} \cite{GrayG1984}. Here, the angles $\theta$ and $\phi$ are the angles of spherical polar coordinates determining the orientation of a particle's main axis, the third angle $\psi$ then specifies how the particle has to be rotated around this axis in order to get into a certain position. If the particle has an axis of continuous rotational symmetry - a typical example for this would be a rod - the axis whose orientation is specified by $\theta$ and $\phi$ can be conveniently chosen to be the symmetry axis. In this case, the angle $\psi$ has no physical relevance and can be ignored, such that the particle's orientation is fully specified by only two angles. Such a particle is referred to as an \textit{uniaxial particle}, a particle without such a symmetry is a \textit{biaxial particle}.

\subsubsection{Orientational ordering tensors}
Whether or not a system of particles is in an ordered phase can be measured using \textit{orientational order parameters}. These can be systematically defined by expanding the orientational distribution function into Cartesian tensors \cite{cremer2012director,te2020relations,te2020orientational}. At second order and for uniaxial particles, this expansion gives rise to the nematic tensor $Q_{ij}$, which is a symmetric traceless tensor. The eigenvalue $S$ of $Q_{ij}$ with the largest absolute value measures the degree of nematic order, the corresponding eigenvector is the nematic director $\vec{n}$ \cite{Andrienko2018}. If the two other eigenvalues are equal to $-S/2$, the system is in an \textit{uniaxial (nematic) phase}. On the other hand, if the system has three distinct eigenvalues, it is in a \textit{biaxial phase} \cite{LuckhurstS2015}.

The distinction between uniaxial and biaxial can thus be made both regarding the particle shapes and regarding the ordered phases. Biaxial particles can also form a uniaxial phase. In principle, uniaxial particles can also form a biaxial phase (if their axes are ordered in such a way that their nematic tensor has three distinct eigenvalues). However, in the absence of additional external influences, phase biaxiality in practice typically only arises from particle biaxiality.

Which order parameters are appropriate, and which ordered phases the system can exhibit, depends on the symmetries of the particles. For example, in a system of square cuboids where the edges have different lengths, one can distinguish a phase in which edges of the same length tend to be parallel from a phase in which edges of the same length are either parallel or orthogonal to each other. For cubes, where all edges have the same length, this distinction would not be meaningful since these phases would be physically indistinguishable.

\begin{figure*}[t]

  \centering

  \includegraphics[width=\linewidth]{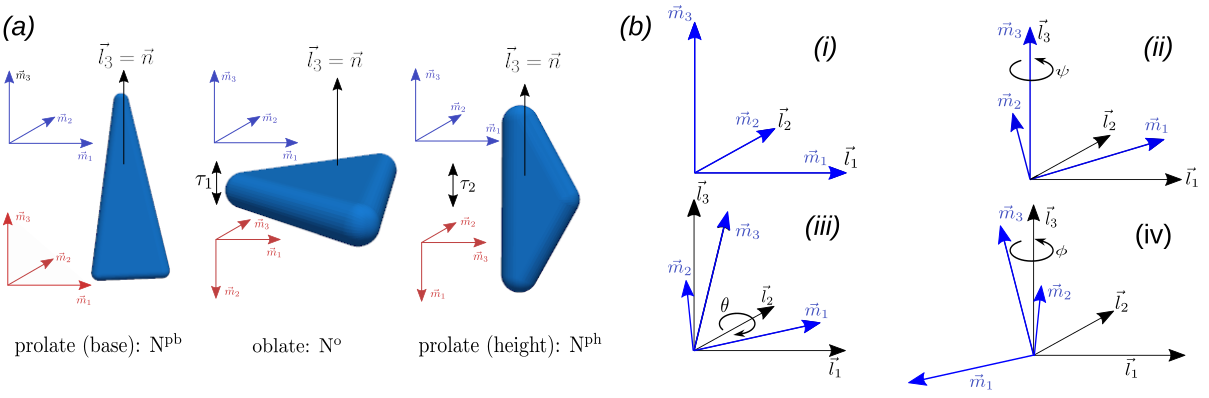}
  \caption{Relation between the body frame
    $(\vec{m}_1,\vec{m}_2,\vec{m}_3)$ and the lab frame $(\vec{l}_1,\vec{l}_2,\vec{l}_3)$.
    \textbf{(a)} Expected alignment of the uniaxial nematic director $\vec{n}$ (specified in the lab frame such that $\vec{l}_3\,||\,\vec{n}$) for different shape ratios $x$ of a spherotriangle, compare Fig.~\ref{fig_bodies}, as indicated by the bottom arrow.
    For small or large $x$, the body has a prolate shape and we expect the height or the base to align with $\vec{n}$, respectively.
    For oblate shapes at intermediate $x$, we expect $\vec{n}$ to be perpendicular to the face of the spherotriangle.
    In each case, the blue coordinates indicate the body frame chosen such that the main particle axis  is parallel to $\vec{m}_3$, while the red coordinates indicate the body frame chosen such that $\vec{m}_3$ always specifies the direction of the height.
    Both conventions result in a different interpretation of the order parameters, which can be identified according to substitutions in Eq.~\eqref{eq_substTAU1} and Eq.~\eqref{eq_substTAU2}, as indicated by the double arrows with corresponding permutation operators $\tau_{1/2}$ annotated.
    Further details are given in the text and Ref.~\onlinecite{rosso2007orientational}.
    In these transformations,  the sign of $\vec{m}_i$ is irrelevant for the set of order parameters considered here.
    \textbf{(b)} Illustration of the three Euler angles, transforming the body frame (blue) to the lab frame (black) via the rotation matrix $\hat{\mathcal{R}}$ from Eq.~\eqref{eq_R}:
    (i) both coordinate frames are initially aligned, in particular $\vec{l}_3=\vec{m}_3$, which is assumed for a proper interpretation of the order parameters defined in Sec.~\ref{sec_orderparametersD2h};
(ii) a rotation by $\psi$ around $\vec{l}_3$ specifies the orientation perpendicular to the body's main axis (irrelevant for uniaxial particles);
(iii) a rotation by $\theta$ around $\vec{l}_2$ specifies the polar orientation of the body's main axis;
(iv) a rotation by $\phi$ around $\vec{l}_3$ specifies the azimuthal orientation of the body's main axis.
    }
    \label{fig_coordinates}
\end{figure*}

This implies that systems of particles whose shape is more complex than that of rods (the most widely studied particle type in liquid crystal physics) can exhibit a much richer phase behavior and require more sophisticated order parameters \cite{te2020relations,rosso2007orientational}. These can also be defined in terms of a systematic expansion of the distribution function in Wigner matrices or (equivalently) Cartesian tensors. Among experimentalists, the Saupe ordering matrix \cite{allender1984landau,LuckhurstS2015}
\begin{equation}
S_{ij}^{\alpha\beta}=\frac{1}{2}\braket{3\,(\vec{m}_i\cdot\vec{l}_\alpha)(\vec{m}_j\cdot\vec{l}_\beta)-\delta_{ij}\delta_{\alpha\beta}}
\label{eq_Saupematrix}
\end{equation}
with the orientational average $\braket{\cdot}$ is particularly popular.
Here, the orthonormal sets $\{\vec{m}_i\}$, $i=1,2,3$ and $\{\vec{l}_\alpha\}$, $\alpha=1,2,3$ constitute a basis fixed in the lab frame and the molecular frame, respectively. The reason one needs a $3 \times 3 \times 3 \times 3$ matrix (namely $S_{ij}^{\alpha\beta}$) in the case of biaxial particles rather than a $3 \times 3$ matrix (namely $Q_{ij}$) for uniaxial particle is that, in the biaxial case, the specification of the particle orientation requires two orthogonal orientation vectors, with the Euler angle $\psi$ specifying the second one. The matrix $S_{ij}^{\alpha\beta}$ roughly corresponds to the second order in the Cartesian expansion of a distribution function depending on all three Euler angles,
see Ref.\ \onlinecite{Turzi2011} for more details.

\subsubsection{Orientational order parameters \label{sec_orderparametersD2h}}

For a particle shape which belongs to the $D_{2\mathrm{h}}$ symmetry group (i.e., with the same symmetry as a cuboid), one can show \cite{rosso2007orientational} that the Saupe ordering matrix has only four independent elements. In this case, it is convenient to work not directly with elements of the Saupe ordering matrix, but with the order parameters
\begin{align}
S&=\frac{3}{2}\braket{(\vec{m}_3\cdot\vec{l}_3)^2 - \frac{1}{3}}\,,\cr
U&=\frac{\sqrt{3}}{2}\braket{(\vec{m}_1\cdot\vec{l}_3)^2-(\vec{m}_2\cdot\vec{l}_3)^2}\,,\cr
P&=\frac{\sqrt{3}}{2}\braket{(\vec{m}_3\cdot\vec{l}_1)^2-(\vec{m}_3\cdot\vec{l}_2)^2}\,,\cr
F&=\frac{1}{2}\braket{(\vec{m}_1\cdot\vec{l}_1)^2-(\vec{m}_1\cdot\vec{l}_2)^2-(\vec{m}_2\cdot\vec{l}_1)^2+(\vec{m}_2\cdot\vec{l}_2)^2}\,,\ \ \cr \label{eq_Xall}
\end{align}
which can be expressed as linear combinations of the four independent matrix elements. The physical meaning of these parameters is as follows:
\begin{itemize}
    \item $S$, the \textit{uniaxial nematic order parameter}, measures whether the axis $\vec{m}_3$ of the molecules is aligned with the lab axis $\vec{l}_3$. If $\vec{m}_3$ is the symmetry axis of a rod and $\vec{l}_3$ is chosen to align with the uniaxial nematic director $\vec{n}$, then $S$ is simply the standard nematic order parameter.
    \item $U$, the \textit{molecular biaxiality order parameter}, measures whether there is molecular biaxiality in a uniaxial phase. If there is a physical difference between the molecular axes $\vec{m}_1$ and $\vec{m}_2$, then it is likely that, on average, there is a difference regarding their probability of being aligned with $\vec{l}_3$.
    \item $P$, the \textit{phase biaxiality order parameter}, measures whether the axis $\vec{m}_3$ is aligned preferably with $\vec{l}_1$ or $\vec{l}_2$. In a perfect uniaxial nematic phase, $\vec{m}_3$ would always be aligned with $\vec{l}_3$. If there are deviations from this alignment, they can either be random ($P=0$) or have a preferred direction. In this scenario (biaxial nematic), there are two preferred axes rather than one. Note that $P$ can be nonzero even if the particles are uniaxial since it only depends on $\vec{m}_3$ (whether that will happen in an actual physical system is another matter).
    \item $F$, the \textit{full
  biaxiality order parameter}, measures whether there is orientational ordering with respect to the axes $\vec{m}_1$ and $\vec{m}_2$. A system with nonzero $F$ is fully biaxial in the sense that both particle shape and phase are biaxial \cite{rosso2007orientational}. If, for example, $\vec{m}_1$ is perfectly aligned with $\vec{l}_1$ and $\vec{m}_2$ with $\vec{l}_2$ (which, since $\vec{l}_1$ and $\vec{l}_2$ can be chosen according to the preferred particle orientations, essentially just means hat the vectors $\vec{m}_1$ of all particles are aligned), then $F$ is maximal. On the other hand, if $\vec{m}_1$ has equal probability of being aligned with $\vec{l}_1$ and $\vec{l}_2$ (uniform distribution of $\vec{m}_1$), then $F$ vanishes.
\end{itemize}

Moreover, we will demonstrate in this work that the order parameters $S$, $U$, $P$ and $F$ can be used not only for particles with $D_{2\mathrm{h}}$ symmetry, but even for certain polar particles.
Cone-like particles, for example, are polar (a mirror reflection at the base changes their physical state).
However, while they may exhibit local polar order in spatially inhomogeneous configurations, they do not form spatially homogeneous polar phases \cite{kubala2023splay}.
Since our focus lies on orientational order in spatially homogeneous phases here, we can, despite the fact that we consider particles with polar symmetries, restrict ourselves to the parameters $S$, $U$, $P$ and $F$.

\subsubsection{Choice of coordinate systems \label{sec_cosy}}

The description of a certain physical state of a system in terms of $S$, $U$, $P$ and $F$ is not unique but it depends on the alignment of both the lab and the body frame, as we will elaborate below and illustrate in Fig.~\ref{fig_coordinates}.
In choosing these coordinate frames, we aim to follow the
 convention that $S$ represents the standard nematic order parameter (the implication of alternative conventions are exemplified in Sec.~\ref{sec_perfectUAorder}).

To fix the coordinate system of the lab frame specified by $\vec{l}_1$, $\vec{l}_2$ and $\vec{l}_3$, we must indicate the direction of the uniaxial director $\vec{n}$.
For example, perfect uniaxial order only corresponds to the case $S=1$ and $P=0$ if we choose $\vec{l}_3$ to align with $\vec{n}$.
Suppose that we instead choose $\vec{l}_2$ to be aligned with $\vec{n}$.
Then we would --- even though the system looks exactly the same --- have $S=-1/2$ and $P=\sqrt{3}/2$ instead of $S=1$ and $P=0$.
To avoid this problem and to ensure that the order parameters have a clear physical interpretation that coincides with the usual one in the uniaxial limit, we always choose the lab frame to be aligned with the \textit{director frame}, i.e., $\vec{l}_3\,||\,\vec{n}$.
It will be shown that in this case the parameter $P$ is irrelevant for describing phase transitions in homogeneous bulk systems.
The parameter $U$, despite being relevant for characterizing the orientational distribution, will turn out not to play a major role for the phase behavior.
This leaves us with the two central order parameters $S$ and $F$, which we use to map out our phase diagrams, where
$S$ measures nematic order in the standard way and a nonzero $F$ then shows that biaxial order is present.

The coordinate system needs to be fixed not only for the lab frame, but also for the body frame, specified by $\vec{m}_1$, $\vec{m}_2$ and $\vec{m}_3$.
For uniaxial particles, the symmetry axis must be chosen to be parallel to $\vec{m}_3$ irrespective of their shape. Otherwise, one would treat the particle as if it was biaxial, since $S$ would not measure the alignment of the symmetry axis to the director and $U$ or $F$ may not vanish.
This choice of the main axis is, however, not as obvious for biaxial particles.
Take, as an example, an isosceles spherotriangle (made of a triangle with two sides of equal length). If the base of such a triangle (the side whose length is different) is much longer than the other two sides, the particles are expected to exhibit nematic order with the director pointing along the base of the triangles. If, on the other hand, the base is sufficiently short, the particles exhibit nematic order with the director pointing in a direction being orthogonal to the base.

Specifically, in this work, we will consider the three different uniaxial nematic phases, which we distinguish by their director orientations assumed for a certain typical particle shape, as depicted in \cref{fig_coordinates} for three representative cases:
(i) a prolate nematic, N$^\text{ph}$, in which the triangles align to the director with their height (when it is much larger than the base length),
(ii) an oblate nematic, N$^\text{o}$, in which the triangles align to the director perpendicular to their face (when height and base are of comparable length such that the particle has an oblate shape),
and, (iii) a prolate nematic, N$^\text{pb}$, in which the triangles align to the director with their base (when it is much larger than their height). We thereby use the words {\textquotedblleft prolate\textquotedblright} and {\textquotedblleft oblate\textquotedblright} for both phases and, as one usually does, for particle shapes. Note that,
similar to the notion of {\textquotedblleft uniaxial\textquotedblright} and {\textquotedblleft biaxial\textquotedblright}, prolate/oblate particle shapes and the emergence of prolate/oblate nematic phases are closely related.
This leaves us with two options for choosing the direction $\vec{m}_3$ (which we identify with the $z$-axis) in the body frame. First, we can choose it parallel to the axis along which the particles presumably align in the case of uniaxial nematic order, which depends on the relative lengths of the different sides of the particle (black coordinate systems drawn in \cref{fig_coordinates}).
This option allows us to follow the convention described above to disregard the order parameter $P$ for each presumed director orientation and is employed in this work.
Second, we can choose to always align $\vec{m}_3$ with, say the height of the triangle
(red coordinate systems drawn in \cref{fig_coordinates}).
This option is more general as there is no need to input assumptions \textit{a priori}, which may impose a bias towards certain director orientations, but it requires to take into account all four order parameters $S$, $U$, $P$ and $F$ without a clear interpretation.

Finally, we note that there exists a well-defined relation between the order parameters obtained from different choices of unit vectors in the body frame, which we summarize below.
Suppose we chose the second option and align the triangle height with the vector $\vec{m}_3$ of the body frame.
If we then want to investigate uniaxial nematic order along the triangle height, N$^\text{ph}$,
 the first option (aligning the particular axis of interest) is trivially equivalent, as indicated in the first panel of \cref{fig_coordinates}.
For uniaxial nematic order along the triangle faces, N$^\text{o}$,
choosing the first option would amount to exchanging the two vectors $\vec{m}_2$ and $\vec{m}_3$,
which defines the action of the permutation operator $\tau_1$ on the basis of the body frame, as indicated in the second panel of \cref{fig_coordinates}.
As detailed in Ref.~\onlinecite{rosso2007orientational}, this procedure allows us to obtain the appropriate order parameters by taking the result of the second option and making the substitutions
\begin{align}
 S&\stackrel{\tau_1}{\rightarrow} -\frac{1}{2}S-\frac{\sqrt{3}}{2}U\,,\ \ \ U\stackrel{\tau_1}{\rightarrow} -\frac{\sqrt{3}}{2}S+\frac{1}{2}U\,, \cr
 P&\stackrel{\tau_1}{\rightarrow} -\frac{1}{2}P-\frac{\sqrt{3}}{2}F\,,\ \ \ F\stackrel{\tau_1}{\rightarrow} -\frac{\sqrt{3}}{2}P+\frac{1}{2}F\,.
 \label{eq_substTAU1}
\end{align}
The same can be done for uniaxial nematic order along the triangle base, N$^\text{pb}$,
where  $\vec{m}_1$ and $\vec{m}_3$  are exchanged by $\tau_2$, as indicated in the third panel of \cref{fig_coordinates}.
The appropriate substitutions are then found to be
\begin{align}
 S&\stackrel{\tau_2}{\rightarrow} -\frac{1}{2}S+\frac{\sqrt{3}}{2}U\,,\ \ \ U\stackrel{\tau_2}{\rightarrow} \frac{\sqrt{3}}{2}S+\frac{1}{2}U\,, \cr
 P&\stackrel{\tau_2}{\rightarrow} -\frac{1}{2}P+\frac{\sqrt{3}}{2}F\,,\ \ \ F\stackrel{\tau_2}{\rightarrow} \frac{\sqrt{3}}{2}P+\frac{1}{2}F\,.
 \label{eq_substTAU2}
\end{align}
An alternative point of view on these substitutions is discussed in Sec.~\ref{sec_perfectUAorder} for a scenario with perfect uniaxial order.

\subsubsection{Representation in terms of Euler angles \label{sec_Eulerangles}}

The definitions of the order parameters \eqref{eq_Xall} are relatively general and require, for practical calculations, a specification of the average $\braket{\cdot}$ that is used.
We assume that the system can be characterized by a distribution $g(\phi, \theta,\psi)$ that gives the probability of finding a particle with an orientation specified by the Euler angles $\phi$, $\theta$ and $\psi$, which connect the body frame to the lab frame.
Here, we follow the convention of Ref.~\onlinecite{rosso2007orientational} and use the $y$-notation.
As illustrated in Fig.~\ref{fig_coordinates}b, the body frame is  first rotated by the angle $\psi\in[0,2\pi]$ around $\vec{l}_3$ (which coincides by default with its $z$-axis, i.e., $\vec{m}_3$).
This is an identity map for uniaxial bodies.
It is followed by a second rotation by the angle $\theta\in[0,\pi]$ around $\vec{l}_2$
and a third one by the angle $\phi\in[0,2\pi]$ around $\vec{l}_3$.
These latter two rotations map the $z$-axis of the body frame onto the unit sphere.

Any vector can be transferred from the body frame to the lab frame via the rotation matrix
\begin{align}
   \hat{\mathcal{R}}:=\hat{R}_3(\phi)\cdot\hat{R}_2(\theta)\cdot\hat{R}_3(\psi)
   \label{eq_R}
\end{align}
accounting for all three Euler angles. Following the convention of using extrinsic rotations, $\hat{R}_\alpha(\gamma)$ is the rotation matrix describing a clockwise rotation by the angle $\gamma$ around the axis of the $\alpha$th unit vector $\vec{l}_\alpha$ of the lab frame.
Explicitly, the components of $\hat{\mathcal{R}}$ read
\begin{align}
  \hat{\mathcal{R}}_{11} &= \cos(\phi)\cos(\theta)\cos(\psi)-\sin(\phi)\sin(\psi)\,, \cr
  \hat{\mathcal{R}}_{12} &= -\sin(\phi)\cos(\psi)-\cos(\phi)\cos(\theta)\sin(\psi)\,, \cr
  \hat{\mathcal{R}}_{13} &= \cos(\phi)\sin(\theta)\,, \cr
  \hat{\mathcal{R}}_{21} &= \cos(\phi)\sin(\psi)+\cos(\theta)\cos(\psi)\sin(\phi)\,, \cr
  \hat{\mathcal{R}}_{22} &= \cos(\phi)\cos(\psi)-\cos(\theta)\sin(\phi)\sin(\psi)\,, \cr
  \hat{\mathcal{R}}_{23} &= \sin(\phi)\sin(\theta)\,, \cr
  \hat{\mathcal{R}}_{31} &= -\cos(\psi)\sin(\theta)\,, \cr
  \hat{\mathcal{R}}_{32} &= \sin(\theta)\sin(\psi)\,, \cr
  \hat{\mathcal{R}}_{33} &= \cos(\theta)\,,
  \label{eq_Rii}
\end{align}
  and these can be conveniently expressed in terms of the basis vectors of the director and body frame as
  \begin{align}
      \hat{\mathcal{R}}_{\alpha i}=\vec{m}_i\cdot\vec{l}_\alpha\,,
      \label{eq_rotlm}
  \end{align}
which allows us to make contact to the order parameters, as defined in Eq.~\eqref{eq_Xall}.

In terms of the Euler angles $(\phi, \theta,\psi)$ and an appropriate orientational distribution $g(\phi, \theta,\psi)$, we can rewrite the four order parameters $S$, $U$, $P$ and $F$ as
\begin{equation}
X=\int \upd\mathbf{O}  \,g(\phi, \theta,\psi)\,f_X(\phi, \theta,\psi) \,,\ \ \ X\in\{S,U,P,F\}\,,
\label{eq_orderparameters}
    \end{equation}
where the angular integral reads
\begin{equation}
\int \upd\mathbf{O}:=\frac{1}{8 \pi^2}\int_{0}^{2\pi}\upd\phi \int_{0}^{\pi}\upd\theta \sin(\theta) \int_{0}^{2\pi}\upd\psi
\end{equation}
and the functions
\begin{align}
    f_S(\phi, \theta,\psi)&=\frac{1}{2}\left(3\cos^2(\theta)-1\right)\,,\label{eq_fS}\\
    f_U(\phi, \theta,\psi)&=\frac{\sqrt{3}}{2}\sin^2(\theta)\cos(2\psi)\,,\label{eq_fU}\\
    f_P(\phi, \theta,\psi)&=\frac{\sqrt{3}}{2}\sin^2(\theta)\cos(2\phi)\,,\label{eq_fP}\\
    f_F(\phi, \theta,\psi)&=\frac{1}{2}\left(1+\cos^2(\theta)\right)\cos(2\phi)\cos(2\psi)\cr
    &\ \ \ \; -\cos(\theta)\sin(2\phi)\sin(2\psi)\,,\label{eq_fF}
\end{align}
are specified by inserting the relation from Eq.~\eqref{eq_rotlm} and the explicit expressions from Eq.~\eqref{eq_Rii} into Eq.~\eqref{eq_Xall}.
In this explicit representation of the order parameters, the choice of coordinate systems discussed in Sec.~\ref{sec_cosy} becomes more intuitive.
The only relevant angle for the uniaxial nematic order parameter~\eqref{eq_fS} is $\theta$.
 For $\theta=0$, i.e., in the case of perfect uniaxial order, $\vec{m}_3$ equals $\vec{l}_3$ (as illustrated in Fig.~\ref{fig_coordinates}).
 Hence the particle axis chosen to align with $\vec{m}_3$ in the body frame sets the uniaxial director, i.e., the maximum of $g$ at $\theta=0$.

\subsection{Fundamental measure theory (FMT)}

\subsubsection{Density functional theory (DFT)}

Based on the celebrated Hohenberg-Kohn theorem \cite{HohenbergK1964} originally developed for quantum-mechanical systems,
classical density functional theory (DFT) \cite{Evans1979} states that there exists a unique functional $\Omega[\rho]$ of the classical number density $\rho$, which gets minimal in equilibrium, i.e., when inserting the equilibrium density $\rho=\rho_{0}$.
Then, the value of the functional equals the grand potential of the system, $\Omega=\Omega[\rho_{0}]$.
For systems of anisotropic particles as considered here, the density $\rho(\bvec{r},\bvec{O})$ depends in general on both positions $\bvec{r}$ and orientations $\bvec{O}$.
Thus, to determine the equilibrium configuration of such a system from a given functional $\Omega[\rho]$, one has to perform a basic minimization by solving the Euler-Lagrange equation
\begin{equation}
    \frac{\delta \Omega[\rho]}{\delta \rho(\bvec{r},\bvec{O})}\bigg|_{\rho=\rho_0}=0\,.
    \label{eq_ELG}
\end{equation}

The general form of the density functional is
\begin{align}
\beta\Omega[\rho]=\int \upd\bvec{r}& \left( \Phi_\text{id}(\bvec{r})+ \Phi_\text{ex}(\bvec{r})\right)\cr
+\int \upd\bvec{r}& \int \upd\mathbf{O}\, \rho(\bvec{r},\bvec{O})(\beta V_\text{ext}(\bvec{r},\bvec{O})-\beta\mu)\,,
\label{eq_DFT}
\end{align}
where $V_\text{ext}(\bvec{r},\bvec{O})$ is an external potential acting on the particles, $\mu$ is their chemical potential and $\beta$ is the inverse temperature.
The free energy density of the system is split into the exactly known contribution
\begin{equation}
    \Phi_\text{id}(\bvec{r})=\int \upd\mathbf{O}\,\rho(\bvec{r},\bvec{O})(\ln(\lambda^3\rho(\bvec{r},\bvec{O}))-1)
    \label{eq_Phiid0}
\end{equation}
of an anisotropic ideal gas with the (irrelevant) thermal volume $\lambda^3$ (comprising trivial integrals over momenta and angular momenta) and the excess free energy density $\Phi_\text{ex}(\bvec{r})$,
which describes the interactions in the system and is not known in general.

\subsubsection{Excess free energy for hard bodies}

 For convex hard bodies, a sophisticated expression for $\Phi_\text{ex}(\bvec{r})$ exists that is based on the seminal work of Rosenfeld, who found an elegant way to decompose the interaction of hard spheres in 1989 \cite{Rosenfeld1989}.
 His FMT functional has been further refined in later work \cite{tarazona2000density,hansen2006WBII,roth2010fundamental,roth2012communication}
 and eventually generalized to more general shapes which can assemble in orientationally ordered phases \cite{hansen2009edFMT,wittmann2015FMMT,wittmann2015THESIS,wittmann2016}.

In the framework of FMT, the hard-body interaction can be conveniently expressed in terms of a set (labeled by the index $\nu$) of weight functions $\omega^{(\nu)}(\bvec{r},\bvec{O})$.
These are local geometrical measures of a hard body $\mathcal{B}$ with surface $\partial\mathcal{B}$.
These functions represent an interacting particle and thus both depend on its position in space and its orientation, as we will specify in Sec.~\ref{sec_homwd}.
This allows us to shift the functional dependence on the density from the excess free energy to a set of weighted densities
\begin{equation}
\label{eq_weighdenorient}
  n_{\nu}(\bvec{r}) =  \int \upd\bvec{r} \int \upd\bvec{O} \,  \rho(\bvec{r}',\bvec{O})\, \omega^{(\nu)}(\bvec{r}-\bvec{r}',\bvec{O}) \,,
\end{equation}
which are defined as an orientationally averaged convolution of the density and the weight functions.

The key feature of FMT is that we can write the excess free energy density
in the general form
\begin{equation}
\Phi_{\text{ex}}
=-n_0\ln(1-n_3)+\frac{\phi_2}{(1-n_3)}+\frac{\phi_3}{(1-n_3)^2}
\label{eq_PhiMM}
  \end{equation}
  solely as a function of weighted densities~\eqref{eq_weighdenorient}.
Here, we use the truncated expansion
  \begin{equation}  \phi_2=n_1n_2-\overrightarrow{n}_1\cdot\overrightarrow{n}_2-\zeta\,\mbox{Tr}\!\left[\overleftrightarrow{n}_1\overleftrightarrow{n}_2\right]
    \label{eq_phi2zeta}
  \end{equation}
involving vectorial and tensorial weighted densities of rank two, the correction factor $\zeta=5/4$ and the Tarazona-Rosenfeld \cite{tarazonarosenfeld1997zero} version of the term
\begin{equation}
\phi_3
=\frac{3}{16\pi}\left(n_2^3-3n_2\mbox{Tr}\!\left[\overleftrightarrow{n}_2^2\right]+2\mbox{Tr}\!\left[\overleftrightarrow{n}_2^3\right]\right)\,.
\label{eq_phi3TR}
  \end{equation}
 Note that a proper choice of the terms $\phi_2$ and $\phi_3$ is particularly important for anisotropic particles, since the standard version of $\phi_3$ for hard spheres, as proposed by Tarazona \cite{tarazona2000density}, leads to qualitatively and quantitatively poor results in the anisotropic case \cite{wittmann2014} (while all versions of Eq.~\eqref{eq_PhiMM} reduce to an appropriate functional for hard spheres if the according weighted densities are used).
On the other hand, more sophisticated choices of both terms are available
  in terms of more complicated geometrical two- and three-body measures
and their appropriate expansions into rotational invariants (in the case of uniaxial particles) \cite{schonhofer2018density}.
We choose to work here with the functional based on Eqs.~\eqref{eq_phi2zeta} and \eqref{eq_phi3TR} for the following reasons:
 It has been demonstrated in detail for hard spherocylinders that the chosen approximation of $\phi_2$ in terms of rank-two tensors does not result in a major offset when locating the isotropic--nematic transition (while it slightly underestimates the difference of coexistence densities).
Most notably, we will demonstrate below that only using rank-two tensors allows us to make direct contact to the general order parameters introduced in Sec.~\ref{sec_SUPF},  which is one of the main goals of the present work.
In doing so, we will gain direct analytic insight for spatially homogeneous systems instead of having to perform a tedious numerical integration, which would be required for other versions of $\Phi_{\text{ex}}$.

\subsubsection{Homogeneous weighted densities \label{sec_homwd}}

In what follows, we are particularly interested in the bulk phase behavior.
Thus, we set
 $V_\text{ext}(\bvec{r},\bvec{O})=0$ in Eq.~\eqref{eq_DFT}
 and focus on spatially homogeneous (but orientationally ordered) phases.
 Hence, we express the orientation-dependent density
 $\rho(\bvec{O})=\rho\, g(\phi, \theta,\psi)$ in terms of the
 homogeneous bulk density $\rho$ and the normalized orientational distribution function  $g(\phi, \theta,\psi)$  entering in Eq.~\eqref{eq_orderparameters}.
With this definition, the free energy density of the ideal gas~\eqref{eq_Phiid0} simplifies to
\begin{equation}
    \Phi_\text{id}=\rho\left(\ln(\lambda^3\rho)-1 + \int \upd\mathbf{O}\,g(\phi, \theta,\psi)\ln g(\phi, \theta,\psi)\right)
    \label{eq_Phiid}
\end{equation}
and the convolution in Eq.~\eqref{eq_weighdenorient}
turns into an ordinary integral over space.
It is thus possible to write all required weighted densities
in an instructive form, which directly incorporates the appropriate orientation dependence of the corresponding weight functions.

The four scalar weighted densities
\begin{align}
    n_{3}&=\rho\int_{\mathcal{B}}\upd\bvec{r}=\rho\, v=\eta\,,\cr
    n_{2}&=\rho\int_{\partial\mathcal{B}}\upd\bvec{r}\,,\cr
    n_{1}&=\rho\int_{\partial\mathcal{B}}\upd\bvec{r} \frac{\mathcal{H}(\bvec{r})}{4\pi}\,,\cr
    n_{0}&=\rho\int_{\partial\mathcal{B}}\upd\bvec{r} \frac{\mathcal{K}(\bvec{r})}{4\pi}=\rho\,
    \label{eq_wdSCALAR}
\end{align}
do not depend on the orientational distribution and represent the particle volume $v$,
its surface area, the surface average of the
local mean curvature $\mathcal{H}(\bvec{r})$ and the surface average of the local Gaussian curvature $\mathcal{K}(\bvec{r})$, respectively.
The latter is equal to one for any (simply connected) hard body.
Moreover, we have defined the packing fraction $\eta$ obtained for $n_{3}$ as the product of the density and the particle volume.
The two vector densities
\begin{align}
   \overrightarrow{n}_{2}&=\rho\int_{\partial\mathcal{B}}\upd\bvec{r}\,\int \upd\mathbf{O}\, g(\phi, \theta,\psi)\,\bvec{n}  \,,\cr
   \overrightarrow{n}_{1}&=\rho\int_{\partial\mathcal{B}}\upd\bvec{r}\,\int \upd\mathbf{O}\, g(\phi, \theta,\psi)\,\frac{\mathcal{H}(\bvec{r})}{4\pi}\bvec{n}\,
    \label{eq_wdVECTOR}
\end{align}
represent averages of the surface unit normal vector $\bvec{n}(\bvec{r},\bvec{O})$.
Finally, the tensor densities
\begin{align}
    \overleftrightarrow{n}_{2}&=\rho\int_{\partial\mathcal{B}}\upd\bvec{r}\,\int \upd\mathbf{O}\,g(\phi, \theta,\psi)\, \bvec{n} \bvec{n}^T\,,\cr
    \overleftrightarrow{n}_{1}&=\rho\int_{\partial\mathcal{B}}\upd\bvec{r}\,\int \upd\mathbf{O}\,g(\phi, \theta,\psi)\,\cr
    &\ \ \ \ \ \ \ \ \ \ \ \ \ \ \ \times\frac{\kappa_{1}(\bvec{r})-\kappa_{2}(\bvec{r})}{8 \pi} \left(\bvec{v}_{1} \bvec{v}_{1}^T -\bvec{v}_{2}  \bvec{v}_{2}^T \right)\,
    \label{eq_wdTENSOR}
\end{align}
are quadratic in the vectors characterizing the particle surface ($\bvec{n} \bvec{n}^T$ denotes a tensor product).
Here, the $\kappa_{i}(\bvec{r})$ denote the local principal curvatures of the surface in the direction of the unit vectors $\bvec{v}_{i}(\bvec{r},\bvec{O})$, $i=1,2$.

Note that the vectorial and tensorial weighted densities represent orientational averages and are therefore crucial for describing orientationally ordered phases.
While the former will always vanish for apolar particle shapes and average to zero for the nematic phases of interest here, the latter will allow us to identify appropriate order parameters for uniaxial and biaxial order.
To show this, we provide below a detailed recipe for how to calculate the weighted densities.

\subsubsection{Explicit calculation of weighted densities: orientational order in FMT \label{sec_wdrecipe}}

To explicitly calculate the weighted densities in Eqs.~\eqref{eq_wdSCALAR}-\eqref{eq_wdTENSOR}, the position dependence in the integrands is to be understood in the sense that $\bvec{r}$ is a coordinate in the body frame
and the orientation dependence of vectors and tensors stems from the relation between the body and the lab frames.
While the volume measure $n_{3}$ always equals the packing fraction,
the surface measures (involving positional integrals over $\partial\mathcal{B}$) can be calculated by a six-step procedure explained in the following. Additional information is provided in appendix~\ref{app_calculationWD}).
To calculate the scalar measures, it is sufficient to follow steps two, three and five, as they do not require an explicit orientational averaging.

The first step is to identify (in the case of uniaxial particles) or choose (in the case of biaxial particles, where this selection may be ambiguous, see Sec.~\ref{sec_cosy}) the primary uniaxial nematic axis for a given particle geometry and ensure that it points in the $z$-direction of the body frame (i.e., the basis vector $\vec{m}_3$).
The second step is to parameterize the surface $\partial\mathcal{B}$ in the body frame using two appropriate parameters.
Due to the additivity of the weight functions, different parts of the surface can be treated independently.
Specifically, for the bodies of interest depicted in Fig.~\ref{fig_bodies}, we need to consider the following two-dimensional manifolds: a cone mantle, disks, portions of a torus (in the limit of circular rings), portions of a sphere, (parts of) cylinder mantles and flat triangles. The corresponding parameterizations are summarized in appendix~\ref{app_parameterization}.
The third step is to calculate all geometrical measures on the surface of each body part using the chosen parameterization in the body frame. This is exemplified for the cone mantle in appendix~\ref{app_measurescone}.
The fourth step is to transfer the surface vectors to the lab frame via the rotation matrix $\hat{\mathcal{R}}$, defined in Eq.~\eqref{eq_R}.
To do so, we calculate $x_i(\bvec{r},\bvec{O})=\hat{\mathcal{R}}_{ij}\bar{x}_j(\bvec{r})$,
where $x_i$ represents the desired $i$th component of $\bvec{n}$, $\bvec{v}_1$ or $\bvec{v}_2$ with full orientation dependence in the lab frame,
$\bar{x}_j$ is the $j$th component of these vectors in the body frame and we have implied summation over the repeated index $j$.

The fifth step is to integrate over the surface of all body parts using the parameterization in appendix~\ref{app_parameterization}.
The sixth and final step is to  perform the orientational average and thereby specify how we can explicitly identify the appropriate order parameters in our density functional.
This can be conveniently achieved by recognizing the relation~\eqref{eq_rotlm} between the components of $\hat{\mathcal{R}}$ and the basis vectors of the two coordinate frames.
In general, the calculation of a tensorial weighted density of rank $k$ (with $k=0$ for scalars and $k=1$ for vectors) will involve  $k$ factors of these components.
For our purpose, we find after some algebra that the average of the squared components can be directly related to the order parameters defined in Eq.~\eqref{eq_Xall}, such that we can set
   \begin{align}
  \left\langle(\hat{\mathcal{R}}_{11})^2\right\rangle&=\,\frac{S-\sqrt{3}P-\sqrt{3}U+3F}{6}+\frac{1}{3}\,, \cr
  \left\langle(\hat{\mathcal{R}}_{12})^2\right\rangle&=\,\frac{S-\sqrt{3}P+\sqrt{3}U-3F}{6}+\frac{1}{3}\,, \cr
  \left\langle(\hat{\mathcal{R}}_{13})^2\right\rangle&=\frac{-S+\sqrt{3}P}{3}+\frac{1}{3}\,, \cr
\left\langle(\hat{\mathcal{R}}_{21})^2\right\rangle&=\,\frac{S+\sqrt{3}P-\sqrt{3}U-3F}{6}+\frac{1}{3}\,, \cr \left\langle(\hat{\mathcal{R}}_{22})^2\right\rangle&=\,\frac{S+\sqrt{3}P+\sqrt{3}U+3F}{6}+\frac{1}{3}\,, \cr
  \left\langle(\hat{\mathcal{R}}_{23})^2\right\rangle&=\frac{-S-\sqrt{3}P}{3}+\frac{1}{3}\,, \cr
  \left\langle(\hat{\mathcal{R}}_{31})^2\right\rangle&=\frac{-S+\sqrt{3}U}{3}+\frac{1}{3}\,, \cr
  \left\langle(\hat{\mathcal{R}}_{32})^2\right\rangle&=\frac{-S-\sqrt{3}U}{3}+\frac{1}{3}\,, \cr
  \left\langle(\hat{\mathcal{R}}_{33})^2\right\rangle&=\frac{2}{3}S+\frac{1}{3}\,,
  \label{eq_Rii2av}
  \end{align}
  while all linear and mixed terms must vanish if no other order parameters do contribute.

The weighted densities calculated from the procedure outlined above are collected in appendix~\ref{app_fullWD} for the bodies depicted in Fig.~\ref{fig_bodies}.
 As expected, the tensorial weighted densities of hard cones and hard cylinders (see appendices~\ref{app_fullWDhcone} and~\ref{app_fullWDhcyl}) do not depend on $U$ and $F$, since their shape is uniaxial, while these order parameters become important for hard isosceles spherotriangles (see appendix~\ref{app_fullWDhst}), whose shape is biaxial.
 A comprehensive account on the order parameters for uniaxial particles is given in appendix~\ref{app_fullWDhsc}, where generalized expressions for the weighted densities of hard spherocylinders are found by taking different limits of the results for spherotriangles.
    In short, we can obtain analytic results for the weighted densities, which can be inserted into Eq.~\eqref{eq_PhiMM}.
    Hence, as a final result, we find the general form
    \begin{equation}
    \Phi_\text{ex}[g]=\Phi_\text{ex}(\rho,S,U,P,F)
    \label{eq_PhiEX_SUPF}
\end{equation}
of the excess free energy density,
which functionally depends on the orientational distribution, as a function of the (homogeneous) density and the four order parameters, which are themselves functionals of the orientational distribution, as they are calculated in FMT according to Eq.~\eqref{eq_orderparameters}.

 Recall from Sec.~\ref{sec_orderparametersD2h} that the number of order parameters may increase when dropping the restriction to particles with a $D_{2\mathrm{h}}$ symmetry.
 However, we argue that isosceles spherotriangles can still be fully described by $S$, $U$, $P$ and $F$ as the only deviation from $D_{2\mathrm{h}}$ symmetry is due to their polar shape, which is not reflected by tensors of even rank.
 In turn, vectorial weighted densities do, in principle, provide additional order parameters measuring polarity,
 but their contribution is found to be negligibly small (see appendix~\ref{app_fullWDhcone} for a more detailed discussion).
This is in line with our expectation that shape polarity is not sufficient stabilize phases with global polar order for hard interactions.
      For lower particle symmetries, such as for arbitrary spherotriangles, we describe in appendix~\ref{app_gentriangle} that we get additional terms proportional to $\langle\hat{\mathcal{R}}_{ij}\hat{\mathcal{R}}_{kl}\rangle$ with $i\neq k$ and/or $j\neq l$ in the tensorial weighted densities of rank two, which represent nondiagonal elements of the Saupe matrix~\eqref{eq_Saupematrix} and can thus not be expressed in terms of our four order parameters alone.
More generally, extending the functional to tensors of higher rank in Eq.~\eqref{eq_phi2zeta} would also give rise to a plethora of additional order parameters.
However, their inclusion would blur the compact analytical picture we wish to provide in the following.

\subsection{Identification of phase transitions}

\subsubsection{Free minimization with all order parameters}

Given the homogeneous ideal gas free energy~\eqref{eq_Phiid} and an excess free energy~\eqref{eq_PhiEX_SUPF} that is a function of the four order parameters $S$, $U$, $P$ and $F$, which can be written as in Eq.~\eqref{eq_orderparameters},
the Euler-Lagrange equation ~\eqref{eq_ELG} becomes
\begin{equation}
    \rho\left(\ln(g(\phi, \theta,\psi))+1\right)+\frac{\delta \Phi_\text{ex}}{\delta g(\phi, \theta,\psi)}=0\,.
\end{equation}
Using the chain rule for the functional derivative, we obtain
\begin{equation}
g(\phi, \theta,\psi)=\mathcal{N}^{-1}\prod_{X}\,e^{\alpha_X^2\,f_X(\phi, \theta,\psi)}\,,
\label{eq_gGEN}
\end{equation}
where $\mathcal{N}$ is a normalization constant that can be determined by the condition
\begin{equation}
1=\int \upd\mathbf{O}\,g(\phi, \theta,\psi)\,
\label{eq_gNORM}
\end{equation}
and where we have defined the intrinsic order parameters
\begin{equation}
\alpha_X^2:=-\frac{1}{\rho}\frac{\partial\Phi_\text{ex}}{\partial X}\,,\ \ \ X\in\{S,U,P,F\}\,.
\label{eq_AGL_GEN}
\end{equation}

Inserting the obtained orientational distribution, Eq.~\eqref{eq_gGEN}, into the definition~\eqref{eq_orderparameters} of the order parameters and solving the five coupled equations~\eqref{eq_gNORM} and \eqref{eq_AGL_GEN} for the five unknowns $\alpha_S$, $\alpha_U$, $\alpha_P$, $\alpha_F$ and $\mathcal{N}$ yields the orientational distribution function of the stable homogeneous phase with full information on biaxiality.
Unfortunately, it is in general not possible to find an explicit expression for $\mathcal{N}$ and thus determine this solution in a closed form.

One option to determine the phase boundaries of a model fluid would be a numerical solution of Eqs.~\eqref{eq_gNORM} and \eqref{eq_AGL_GEN},
where the global minimum could be identified from comparing the corresponding free energy $\Phi=\Phi_\text{id}+\Phi_\text{ex}$ (evaluated for the obtained orientational distributions) in the case of multiple solutions.
Moreover, as discussed in Sec.~\ref{sec_cosy}, a minimization with respect to all four order parameters
   would result in a multitude of physically equivalent solutions due to spontaneous symmetry breaking.
  Instead, our goal is to get analytic insight and, in particular, also understand which are the most relevant order parameters for characterizing the different phases.
        Thus we proceed step by step and first investigate in Sec.~\ref{sec_uniaxial}
        the scenario of uniaxial nematic order, solely characterized by $S$,
         before considering in Secs.~\ref{sec_perfectUAorder} and~\ref{sec_perturbedUAorder} the role of the other order parameters $Y\in\{U,P,F\}$ with respect to this reference state.

\subsubsection{Phase transitions involving uniaxial order \label{sec_uniaxial}}

For uniaxial particles and an appropriate choice of the coordinate systems (in which both the director and the symmetry axis point in the $z$-direction of the laboratory and molecular frame, respectively), the only order parameter relevant for the homogeneous bulk phase behavior is $S$.
Thus, we may assume $\Phi_\text{ex}[g]=\Phi_\text{ex}(\rho,S)$ instead of the general form \eqref{eq_PhiEX_SUPF}.
We are left with only two equations (Eq.~\eqref{eq_gNORM} and Eq.~\eqref{eq_AGL_GEN} for $X=S$), which can be solved explicitly \cite{hansen2010tensorial,wittmann2014}, as we briefly recapitulate below.

Let us first define $\alpha^2:=3\alpha_S^2/2$ for later notational convenience, such that Eq.~\eqref{eq_AGL_GEN} becomes
\begin{equation}
\alpha^2:=-\frac{2}{3\rho}\frac{\partial\Phi_\text{ex}}{\partial S}\,.
\label{eq_AGL_GEN_Salpha}
\end{equation}

Normalization of the orientational distribution~\eqref{eq_gGEN} with $\alpha_U=\alpha_P=\alpha_F=0$ yields the explicit expression
\begin{equation}
g=g(\alpha,\cos\theta)=\frac{\alpha}{\mathcal{D}(\alpha)}  \exp\!\left(-\alpha^2\left(1- \cos^2\theta\right) \right)
\label{eq_gUNI}
\end{equation}
with Dawson's integral
\begin{equation}
 \mathcal{D}(\alpha)= \exp(-\alpha^2)\int_0^{\alpha}\mathrm{d}u\exp(u^2) \,.
 \end{equation}
Inserting Eq.~\eqref{eq_gUNI} into Eq.~\eqref{eq_orderparameters}, the nematic order parameter
\begin{eqnarray}
 S(\alpha)=\frac{3}{4\alpha\mathcal{D}(\alpha)}-\frac{3}{4\alpha^2}-\frac{1}{2}
\label{eq_Sdef}
\end{eqnarray}
is obtained as a function of $\alpha$.
With this result, we are left with the task to solve a single equation, Eq.~\eqref{eq_AGL_GEN_Salpha}, in a self-consistent way to obtain the solution $\alpha$ that minimizes the free energy at a given density $\rho$.
Vice versa, the solution for the density as a function of $\alpha$ can even be found in a closed analytic form.

To determine the densities of two coexisting phases we need to impose equilibrium conditions. As the temperature $T$ only enters as a scaling factor that has the same value everywhere in the system, thermal equilibrium is always ensured. However, we need to demand that the two phases are in chemical and mechanical equilibrium.
Hence, the chemical potential
\begin{align}
    \beta\mu=\frac{\partial(\Phi_\text{id}+\Phi_\text{ex})}{\partial\rho}
\end{align}
and the pressure
\begin{align}
    \beta p=\beta \mu \rho - (\Phi_\text{id}+\Phi_\text{ex})
\end{align}
 must be equal.
The phases that can be compared in this way are
the isotropic phase, characterized by the absence of any orientational order ($\alpha=0$)
and (uniaxial) nematic phases found as the nontrivial solution $\alpha$ of Eq.~\eqref{eq_AGL_GEN_Salpha} for a given functional.

For particles with biaxial shape, let us recall that, by choosing different particle axes which may align with the director,
we can also examine coexistence between different uniaxial nematic phases in this way (assuming that no other order parameters are relevant).

\subsubsection{Biaxiality for perfect uniaxial order \label{sec_perfectUAorder}}

A simple way to investigate the uniaxial--biaxial transition is to assume perfect uniaxial order by fixing the main particle axis in space.
This approximation reduces the complexity of the problem  to that of locating the isotropic--nematic transition in two spatial dimensions.
To make this apparent, we align the desired axis in the $\vec{m}_3$-direction of the molecular frame (first option discussed in Sec.~\ref{sec_cosy} and blue coordinate frames in Fig.~\ref{fig_coordinates})
and then set $\theta=\psi=0$ in Eqs.~\eqref{eq_fS}-\eqref{eq_fF}, such that the orientational order (perpendicular to the main axis) is characterized solely by the remaining polar angle $\phi$.
Doing so, we directly find that
\begin{equation}
S=1\,,\ U=P=0\,,\ F=S_{2d}\,,
\label{eq_OPsPA}
\end{equation}
 where
 \begin{equation}
S_{2d}=\frac{1}{2\pi}\int \upd\phi \,g_{2d}(\phi)\,\cos(2\phi) \,
\label{eq_orderparameter2d}
    \end{equation}
is a two-dimensional nematic order parameter and $g_{2d}(\phi)$ the according orientational distribution.
The remaining analysis is analogous to that in Sec.~\ref{sec_uniaxial} and has been performed in Ref.~\onlinecite{wittmann2017twodim} for two-dimensional rods, where it was demonstrated that the resulting self-consistency equation only has one stable solution.
Hence, the uniaxial--biaxial transition density can be identified in a closed analytic form when using the approximation of perfect uniaxial order.

Before proceeding in Sec.~\ref{sec_perturbedUAorder} with a more accurate method to identify the onset of biaxial order, let us revisit alternative choices to align a biaxial particle in the body frame.
As in the final paragraph of Sec.~\ref{sec_cosy}, let us focus on an isosceles spherotriangle and suppose that we always align the triangle height with $\vec{m}_3$.
Now, in the special case of perfect uniaxial order, the manipulations necessary to describe a deviating director alignment can be intuitively illustrated by simply rotating the body frame instead of redefining its coordinates.
Specifically, to achieve a perfect alignment of the director $\vec{n}$ perpendicular to the triangle face ($\vec{m}_2\,||\,\vec{n}$, compare the second case in Fig.~\ref{fig_coordinates}a), we can rotate the body frame by $\psi=\pi/2$, $\theta=\pi/2$ and consider a shifted polar angle $\phi\rightarrow\phi+\pi/2$ in Eqs.~\eqref{eq_fS}-\eqref{eq_fF}, such that $S=-1/2$, $U=-\sqrt{3}/2$, $P=\sqrt{3}S_{2d}/2$ and $F=-S_{2d}/2$.
    This result is equivalent to making the substitutions in Eq.~\eqref{eq_substTAU1} and then choosing the order parameters according to Eq.~\eqref{eq_OPsPA}.
   Moreover, a director perfectly aligned with the baseline of the triangle ($\vec{m}_1\,||\,\vec{n}$, compare the third case in Fig.~\ref{fig_coordinates}a) corresponds to $\psi=0$, $\theta=\pi/2$ and a free $\phi$ in Eqs.~\eqref{eq_fS}-\eqref{eq_fF}, such that $S=-1/2$, $U=\sqrt{3}/2$, $P=\sqrt{3}S_{2d}/2$ and $F=S_{2d}/2$.
    This result is equivalent to making the substitutions in Eq.~\eqref{eq_substTAU2} and then using Eq.~\eqref{eq_OPsPA}.

In summary, the above examples allow us to make sense of the altered interpretation of the order parameters upon dropping the convention that $\vec{m}_3$ denotes the particle axis which preferably aligned with the uniaxial director $\vec{n}$:
in general, even perfect uniaxial order ($S_{2d}=0$) cannot be described by $S$ alone.
However, the common convention, used in Sec.~\ref{sec_uniaxial}, that the order parameter $S$ measures the degree of uniaxial alignment of the main particle axis (chosen parallel to $\vec{m}_3$) with the director (chosen parallel to $\vec{l}_3$), can always be restored by a redefinition of the body frame.

\subsubsection{Biaxiality as perturbation of uniaxial order \label{sec_perturbedUAorder}}

While the assumption of perfectly uniaxial order made in Sec.~\ref{sec_perfectUAorder} is helpful to get a feeling for the relevant order parameters describing biaxiality, it may be a very crude approximation in practice.
A more reliable calculation of the uniaxial--biaxial transition is by investigating the instability of the uniaxial solution, Eq.~\eqref{eq_gUNI}, under small perturbations related to the order parameter $F$.
This strategy will allow us for an exact location of the transition under the two assumptions that it is of second order (and can thus be identified as the limit of stability of the uniaxial phase) and that no other order parameter is relevant.
To assess the latter assumption, we introduce for the sake of generality in the following presentation a dummy parameter $Y\in\{U,P,F\}$, representing any of the three remaining order parameters
and consider it as a perturbation to a phase with $S$ being the only nonzero order parameter,
which we refer to in what follows as the \textit{simple uniaxial phase}.

Assuming that there is only one relevant order parameter $Y$ in addition to $S$,
let us first simplify the general  orientational distribution~\eqref{eq_gGEN}  to (recalling that $\alpha_S^2=2\alpha^2/3$)
\begin{align}
g(\phi, \theta,\psi)&=\mathcal{N}^{-1}g_0(\phi, \theta,\psi)\,,\\
g_0(\phi, \theta,\psi)&=e^{\frac{2\alpha^2}{3}\,f_S(\phi, \theta,\psi)+\alpha_Y^2\,f_Y(\phi, \theta,\psi)}\,,\label{eq_g0Y}\\
\mathcal{N}&=\int \upd\mathbf{O}\,g_0(\phi, \theta,\psi)\label{eq_normdef}\,.
\end{align}
Then, the free energy density \eqref{eq_Phiid} of the ideal gas can be rewritten in the explicit form
\begin{align}
\!\!\!\!\Phi_\text{id}=\rho\left(-\ln\mathcal{N}+\frac{2\alpha^2}{3}\,S+\alpha_Y^2\,Y+\ln(\rho\Lambda^3)-1\right)\,,
\label{eq_PhiID1}
\end{align}
where $\mathcal{N}$, $S$ and $Y$ are, in general, yet unknown functions of $\alpha$ and $\alpha_Y$,
compare Eq.~\eqref{eq_gUNI} and Eq.~\eqref{eq_Sdef} in the special case $\alpha_Y=0$.
 Therefore, the total free energy density $\Phi(\alpha,\alpha_Y,S,Y,\mathcal{N})$ depends on $\alpha$ and $\alpha_Y$ both explicitly and implicitly through $\mathcal{N}(\alpha,\alpha_Y)$, $S(\alpha,\alpha_Y)$ and $Y(\alpha,\alpha_Y)$.

To investigate the stability of a uniaxial solution against perturbations due to the order parameter $Y$, we take a look at the minimum of $\Phi$ as a function of $\alpha_Y^2$, assuming that the value of $\alpha$, at a given density $\rho$, is not affected by $\alpha_Y$ at first order.
For the ideal term~\eqref{eq_PhiID1}, we get
\begin{align}
\frac{\mathrm{d}\Phi_\text{id}}{\mathrm{d} \alpha_Y^2}&=\rho\left(-\frac{1}{\mathcal{N}}\frac{\partial\mathcal{N}}{\partial \alpha_Y^2}+\frac{2\alpha^2}{3}\,\frac{\partial S}{\partial \alpha_Y^2}+Y+\alpha_Y^2\,\frac{\partial Y}{\partial \alpha_Y^2}\right)\!\!\!\!\!\!\!\!\cr
&=\rho\left(\frac{2\alpha^2}{3}\,\frac{\partial S}{\partial \alpha_Y^2}+\alpha_Y^2\,\frac{\partial Y}{\partial \alpha_Y^2}\right)\,,
\label{eq_firstderPHIid}
\end{align}
where we have used that the first term in the first line is equal to $-Y$, which can be verified by inserting the definition of $\mathcal{N}$ from Eq.~\eqref{eq_normdef} with $g_0$ from Eq.~\eqref{eq_g0Y} and calculating the derivative within the integral.
As we consider a perturbation of a simple uniaxial phase, we insert the equilibrium condition from Eq.~\eqref{eq_AGL_GEN_Salpha}, which must hold for $\alpha_Y=0$,
such that the derivative of the excess term~\eqref{eq_PhiEX_SUPF} becomes
\begin{align}
\frac{\mathrm{d}\Phi_\text{ex}}{\mathrm{d} \alpha_Y^2}&=\frac{\partial\Phi_\text{ex}}{\partial Y}\frac{\partial Y}{\partial \alpha_Y^2}-\rho\,\frac{2\alpha^2}{3}\,\frac{\partial S}{\partial \alpha_Y^2}\,.
\label{eq_firstderPHIex}
\end{align}
Adding up Eqs.~\eqref{eq_firstderPHIid} and~\eqref{eq_firstderPHIex}, we get
\begin{align}
\frac{\mathrm{d}\Phi}{\mathrm{d} \alpha_Y^2}&=\left(\rho\alpha_Y^2+\frac{\partial\Phi_\text{ex}}{\partial Y}\right)\frac{\partial Y}{\partial \alpha_Y^2}\,.
\label{eq_firstderPHI}
\end{align}
Since the order parameter $Y$ is a monotonous function of $\alpha_Y$ (both increase for increasing order), i.e., $\partial Y/\partial \alpha_Y^2>0$, the extremal condition $\mathrm{d}\Phi/\mathrm{d} \alpha_Y^2=0$ consistently recovers Eq.~\eqref{eq_AGL_GEN} for $X=Y$.
This condition and the one for $X=S$, i.e., Eq.~\eqref{eq_AGL_GEN_Salpha}, must be mutually fulfilled in a stable system.
In particular, the stability of the simple uniaxial solution has the necessary condition
\begin{align}
\mathcal{E}_Y(\alpha):=\left(\frac{\partial\Phi_\text{ex}}{\partial Y}\frac{\partial Y}{\partial \alpha_Y^2}
\right)_{\alpha_Y=0}=0\,.
\label{eq_mincondYfd}
\end{align}
If it does not hold, the simple uniaxial phase is not stable for the given value of $\alpha$ and we anticipate that $\alpha_Y>0$.

What is left to be determined in the case that Eq.~\eqref{eq_mincondYfd} is fulfilled is whether the corresponding extremal point of the free energy in Eq.~\eqref{eq_firstderPHI}, is indeed a minimum, which is a sufficient condition for the stability of the simple uniaxial solution with $\alpha_Y=0$ against additional order associated with the parameter $Y\in\{U,P,F\}$.
 To be able to answer this question, we must calculate the second derivative of $\Phi$ with respect to $\alpha_Y^2$ and determine its sign.
To do so, we take another derivative of Eq.~\eqref{eq_firstderPHI}, insert Eq.~\eqref{eq_AGL_GEN} for $X=Y$ (which removes the second derivative of $Y$), divide by $\rho$ and divide by $\partial Y/\partial \alpha_Y^2$.
This yields the condition
\begin{align}
\mathcal{A}_Y(\alpha):=1+\left(\frac{\mathrm{d}}{\mathrm{d} \alpha_Y^2}\frac{1}{\rho}\frac{\partial\Phi_\text{ex}}{\partial Y}
\right)_{\alpha_Y=0}=0\,,
\label{eq_mincondY}
\end{align}
for the value of $\alpha$, as determined from solving Eq.~\eqref{eq_AGL_GEN_Salpha}, at which the simple uniaxial phase becomes unstable if the free energy has an extremal point at $Y=\alpha_Y=0$.
Then, the values of $\alpha$ for which $\mathcal{A}_Y(\alpha)>0$ holds indicate the stability range of the simple uniaxial solution (if $\mathcal{E}_Y(\alpha)=0$ and the value of $\alpha$ is beyond isotropic--uniaxial coexistence).

Taking a closer look at the stability function in Eq.~\eqref{eq_mincondY}, we realize that its evaluation only requires the knowledge of all order parameters up to the first order in $\alpha_Y^2$.
This result can be obtained analytically by expanding Eq.~\eqref{eq_g0Y} according to
\begin{align}
g_0(\phi, \theta,\psi)&=e^{\frac{2\alpha^2}{3}\,f_S(\phi, \theta,\psi)}(1+\alpha_Y^2\,f_Y(\phi, \theta,\psi))+\mathcal{O}(\alpha_Y^4)\,\label{eq_g0Yexp}
\end{align}
and performing an explicit normalization according to Eq.~\eqref{eq_normdef}.
Truncating the expansion of $g_0$ after the term of order $\alpha_Y^2$, we find that the nematic order parameter is unaffected for all $Y\in\{U,P,F\}$, i.e., we have
\begin{align}
S(\alpha,\alpha_Y)=S(\alpha)+\mathcal{O}(\alpha_Y^4)\,
\label{eq_SalphaY}
\end{align}
with $S(\alpha)$ given by Eq.~\eqref{eq_Sdef},
while we obtain the leading terms
\begin{align}
\!\!\!\!U(\alpha,\alpha_U)&=\frac{(-3-2\alpha^2)S(\alpha)+2\alpha^2}{8\alpha^2}\,\alpha_U^2+\mathcal{O}(\alpha_U^4)\,,\label{eq_UalphaY}\\
\!\!\!\!P(\alpha,\alpha_P)&=\frac{(-3-2\alpha^2)S(\alpha)+2\alpha^2}{8\alpha^2}\,\alpha_P^2+\mathcal{O}(\alpha_P^4)\,,\\
\!\!\!\!F(\alpha,\alpha_F)&=\frac{(-3+14\alpha^2)S(\alpha)+10\alpha^2}{48\alpha^2}\,\alpha_F^2+\mathcal{O}(\alpha_F^4)\,
\label{eq_FalphaY}
\end{align}
of the other order parameters.
 In each case, the coefficient of $\alpha_Y^2$ is positive for all finite values of $\alpha$, justifying the assumption $\partial Y/\partial \alpha_Y^2>0$ in deriving Eq.~\eqref{eq_mincondY}.
Putting everything together, an analytic expression for $\mathcal{A}_Y(\alpha)$ in Eq.~\eqref{eq_mincondY} with $Y\in\{U,P,F\}$ can be explicitly found by inserting, (i), the explicit uniaxial solution $\rho(\alpha)$ of Eq.~\eqref{eq_AGL_GEN_Salpha} with $Y=0$,
(ii), $S(\alpha,\alpha_Y)$ from Eq.~\eqref{eq_SalphaY}, where only the leading term is relevant,
and, (iii), either of Eqs.~\eqref{eq_UalphaY}-\eqref{eq_FalphaY}.

The whole calculation is analogous for the general orientational distribution from Eq.~\eqref{eq_gGEN}, where all four order parameters are taken into account simultaneously.
In this case, all generalized expansions~\eqref{eq_SalphaY}-\eqref{eq_FalphaY} would contain constant terms in $\alpha$, reflecting the rotational invariance of the problem.
This symmetry is broken by choosing appropriately aligned coordinate frames.
In particular, for our common choice described in Sec.~\ref{sec_cosy},
we  argue that we can assume $P=0$ in our perturbation analysis without loss of generality.

\section{Results for ordering behavior \label{sec_results}}

Next we corroborate the theoretical conclusions drawn in Sec.~\ref{sec_theory} by
 investigating the ordering behavior of the hard bodies illustrated in Fig.~\ref{fig_bodies}.
As uniaxial shapes we consider a cylinder (left) with diameter $D$ and height $H$ (which equals its mantle length $L$), as well as a cone (middle) with circular base area of diameter $D$ and height $H$.
 To define in each case an aspect ratio $l$ measuring the particle anisotropy,
 we choose the convention $l=H/D$.
As a biaxial shape we consider an isosceles spherotriangle (right), defined as the parallel set at distance $D/2$ of an isosceles triangle with base length $A$ and two side lengths $B$ (or, in other words, a triangular prism whose three side faces are capped by cylindrical halves connected by spherical parts).
 For the spherotriangle, we define the aspect ratio $l=(A+2B)/(2D)$ such that it is consistent with the typical convention $l=L/D$ for a spherocylinder with cylindrical mantle length $L$ and diameter $D$.
 Specifically, the spherocylindrical shape is recovered in two limits: upon setting either $A=L$ and $B=L/2$ or $A=0$ and $B=L$.
 Hence, in addition to $l$, we need a second dimensionless parameter to fully describe the shape of a spherotriangle.
 By defining the shape ratio $x$ as
 \begin{align}
     x=\frac{A}{2B}=\sin\gamma\,,
 \end{align}
we describe isosceles triangles of all possible opening angles $2\gamma$,
 which reduce to  spherocylinders in the limiting cases  $x=0$ and $x=1$.
 The corresponding weighted densities required to construct the functional~\eqref{eq_PhiEX_SUPF} are stated in appendix~\ref{app_fullWD} for each particle shape.

Our general strategy is to first justify in Sec.~\ref{sec_RESop} that the order parameters $U$ and $P$ can be disregarded in our treatment of the homogeneous bulk phases, where we relate the onset of  biaxial order to a nonzero value of $F$.
Then, we discuss the phase diagrams of the different hard-body fluids, where we only focus on homogeneous phases.
In doing so we neglect the expected transition to positionally ordered liquid crystal phases and the solid state.
Therefore, we draw our phase diagrams in Figs.~\ref{fig_conecyl} and~\ref{fig_PD}
with backgrounds fading into gray for an increasing packing fraction, indicating the increasing probability that the presented states are only metastable, as other phases might be predicted by the functional.
In Sec.~\ref{sec_RESuni}, we compare the isotropic--nematic transition of two uniaxial shapes which possess the same limiting behavior for extreme aspect ratios: a polar cylinder and an apolar cone.
In Sec.~\ref{sec_RESbi}, we turn to biaxial isosceles spherotriangles.
The onset of biaxial order is discussed further in appendix~\ref{app_PA}.

\begin{figure}[t]
    \centering
    \includegraphics[width=0.45\textwidth]{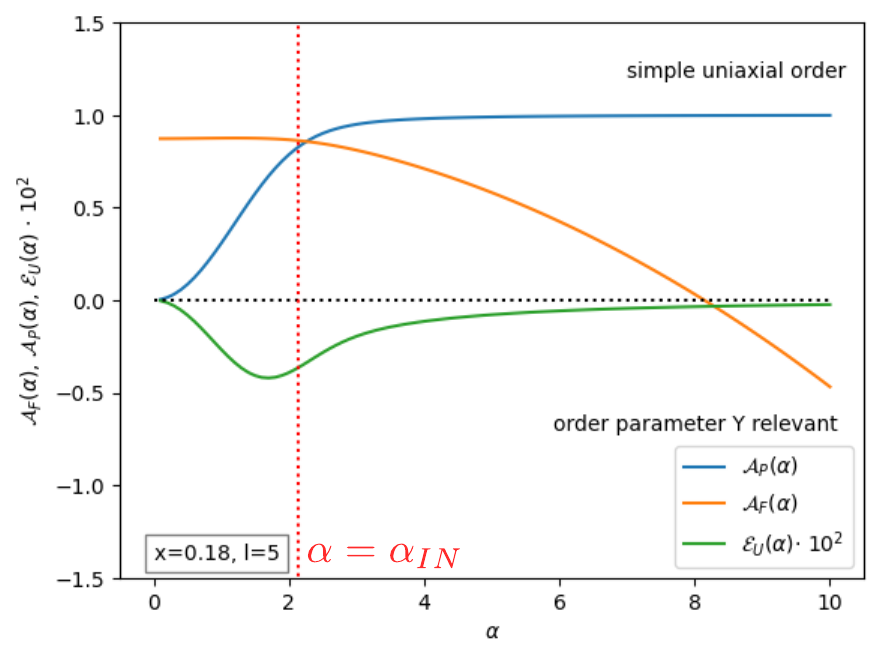}
    \caption{Stability of the uniaxial nematic phase under different perturbations, investigated here for hard spherotriangles with $x=0.18$ and $l=5$.
     The results are representative for all dominant director orientations at all shape ratios considered, compare Fig.~\ref{fig_PD}.
   According to the legend, we show $\mathcal{E}_U(\alpha)$ from Eq.~\eqref{eq_mincondYfd} and $\mathcal{A}_P(\alpha)$ and $\mathcal{A}_F(\alpha)$ from Eq.~\eqref{eq_mincondY} as a function of $\alpha$
   (not shown are the trivial results $\mathcal{E}_P(\alpha)=\mathcal{E}_F(\alpha)=0$ and the meaningless function $\mathcal{A}_U(\alpha)$).
   As annotated, simple uniaxial order, solely specified by $\alpha$ and thus $S$, is stable against a perturbation in $Y\in\{U,P,F\}$ if $\mathcal{E}_Y(\alpha)=0$ and $\mathcal{A}_Y(\alpha)>0$, where the horizontal black dotted line serves as a guide to the eye.
       Hence, the value of $\alpha$ for which $\mathcal{A}_F(\alpha)=0$ indicates the uniaxial--biaxial transition.
    This stability analysis of uniaxial order is only meaningful beyond coexistence with the isotropic phase for $\alpha\geq\alpha_\text{IU}$, as indicated by the
     dotted vertical line.
     }
    \label{fig_perturbation}
\end{figure}

\subsection{Relevant order parameters \label{sec_RESop}}

To illustrate the relevance of the different order parameters,
we perform the perturbation analysis of simple uniaxial nematic order, as described in
Sec.~\ref{sec_perturbedUAorder}, to identify whether an appropriate description of the state point of interest requires to account for the order parameters $U$, $P$ or $F$ in addition to the standard nematic order parameter $S$.
We choose to work here with an isosceles spherotriangle (whose height specifies the main axis) since it is
the most general shape considered in our investigation and because its weighted densities in appendix~\ref{app_fullWDhst} depend on all four order parameters.
Moreover, this biaxial shape reduces to a uniaxial spherocylinder when either the length of the base line becomes zero ($x=0$) or its height vanishes ($x=1$).
The results of both limits differ due to the different direction of the assumed symmetry axis within the body frame, as further discussed in appendix~\ref{app_fullWDhsc}.
In the first case, the only remaining order parameters are $S$ and $P$, as expected for uniaxial particles, while, in the second case, the spherocylinder is formally treated as being a biaxial particle.

As a first step of our perturbation analysis, we must check the first derivatives of the free energy for the simple uniaxial reference case $\alpha_Y=0$ with $Y\in\{U,P,F\}$ according to Eq.~\eqref{eq_mincondYfd}.
Indeed, we find that the free energy has an extremal point for $\alpha_P=0$ and $\alpha_F=0$, i.e., $\mathcal{E}_P(\alpha)=0$ and $\mathcal{E}_F(\alpha)=0$ for all values of $\alpha$.
Regarding a perturbation in terms of the molecular biaxiality order parameter $U$, the free energy has a negative slope at $\alpha_U=0$, i.e., $\mathcal{E}_U(\alpha)<0$ for $0<\alpha<\infty$, as shown in Fig.~\ref{fig_perturbation}.
This suggests that $U$ affects all ordered phases of biaxial particles.
 Accordingly, it was shown in Ref.~\onlinecite{cuetos2017phase} by minimizing a modified Onsager functional with respect to a trial orientational distribution  that  (a parameter closely related to) $U$ is nonzero at the isotropic--uniaxial coexistence of hard cuboids.
 However, ignoring this parameter affects the calculated transition densities  only marginally \cite{cuetos2017phase}.
Moreover, the decreasing absolute value $|\mathcal{E}_U(\alpha)|$ of the slope for large $\alpha$ in Fig.~\ref{fig_perturbation} suggests that the effect of $U$ should become smaller and smaller with increasingly strong uniaxial order and eventually turn fully irrelevant in the limit $\alpha\rightarrow\infty$ (or $S\rightarrow1$), which is consistent with the prediction $U\rightarrow0$ in Eq.~\eqref{eq_OPsPA}.
As biaxial phases imply a large degree of uniaxial order and because $U$ does not measure phase biaxiality, we assume that $U=0$ as a presumably good approximation for making analytic progress.

As a second step, we take a closer look at
the stability functions~\eqref{eq_mincondY} $\mathcal{A}_Y(\alpha)$ with $Y\in\{P,F\}$ ($\mathcal{A}_U(\alpha)$ is only meaningful at $\alpha=0$, since we found that there is otherwise no extremal point of the free energy at $\alpha_U=0$).
As detailed in Sec.~\ref{sec_perturbedUAorder}, simple uniaxial order, where only $S$ has a nonzero value, does not appropriately characterize the system if $\mathcal{A}_Y(\alpha)<0$, which indicates that the stable state should have a nonzero value of $Y$.
Representative results for the stability functions are shown in Fig.~\ref{fig_perturbation}.
 In general, $\mathcal{A}_P$ increases with increasing $\alpha$,
while the opposite trend is observed for $\mathcal{A}_F$.
Again, the limit $\alpha\rightarrow\infty$ (or $S\rightarrow1$) of perfect uniaxial order can be directly understood from the order parameters in Eq.~\eqref{eq_OPsPA}.
As $P\rightarrow0$, we always find that $\mathcal{A}_P\rightarrow1$, because then the term in Eq.~\eqref{eq_mincondY} involving the derivatives of $\Phi_\text{ex}$ does not contribute.
In contrast, the limiting behavior of $\mathcal{A}_F$ depends on the particular system, since $F$ does not vanish.

 Most importantly, Fig.~\ref{fig_perturbation} reveals that the central order parameter to characterize phase biaxiality in our approach is $F$ and we can use the criterion $\mathcal{A}_F(\alpha)=0$ to predict the onset of biaxial order.
Indeed, $\mathcal{A}_F(\alpha)$ is generally found to change its sign at values of $\alpha$ that are larger than $\alpha_\text{IU}$ at the isotropic--uniaxial transition.
In contrast, we find $\mathcal{A}_P(\alpha)>0$ for all $\alpha>0$, which implies that the simple uniaxial phase is always stable against perturbations in $P$, such that the order parameter $P$ is not relevant for identifying the onset of biaxial order in our setup.
 Only within the biaxial phase, we find by generalizing Eq.~\eqref{eq_mincondYfd} that $P$ can take nonzero values, because $\partial\Phi_\text{ex}/\partial P<0$ at $P=0$  does no longer vanish if $F>0$, but this does not affect the phase boundaries.
 We also note that the result $\mathcal{A}_P(0)=0$ suggests that the isotropic phase can equally be destabilized by a nonzero value of $P$ instead of $S$ (which corresponds to a different director alignment in the lab frame), but neither by $U$ nor $F$.

\subsection{Uniaxial bodies: effect of shape polarity \label{sec_RESuni}}

\begin{figure}[t]
    \centering
    \includegraphics[width=0.45\textwidth]{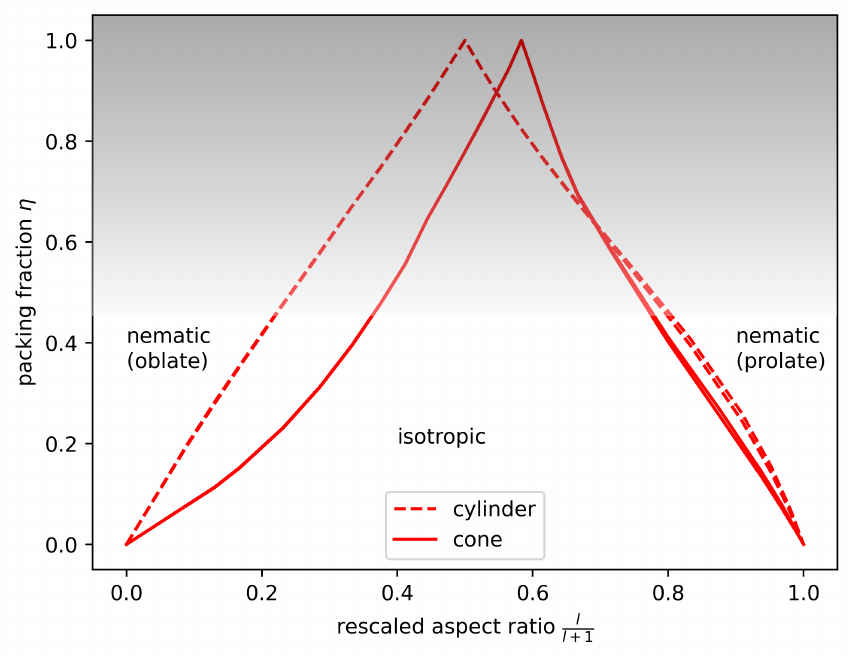}
    \caption{ Isotropic--nematic transition of hard cones (solid lines) and hard cylinders (dotted lines) depending on the packing fraction $\eta$ and the rescaled aspect ratio $l/(1+l)$.
    The transition is always found to be of first order, but the density difference between coexisting phases is not always visible.
    The limits $l/(1+l)=0$ and $l/(1+l)=1$ correspond to hard disks (oblate nematic phase) and rods (prolate nematic phase), respectively.
    As detailed at the beginning of Sec.~\ref{sec_results}, our functional is expected to predict phases with positional order at larger packing fractions.
    Hence, we indicate the region $\eta\gtrsim0.45$ of the phase diagram, where positional order is expected when applying the used functional to hard spherocylinders \cite{wittmann2014},  by shading the background in light gray.}
    \label{fig_conecyl}
\end{figure}

\begin{figure*}[t]
    \centering
    \includegraphics[width=0.485\textwidth]{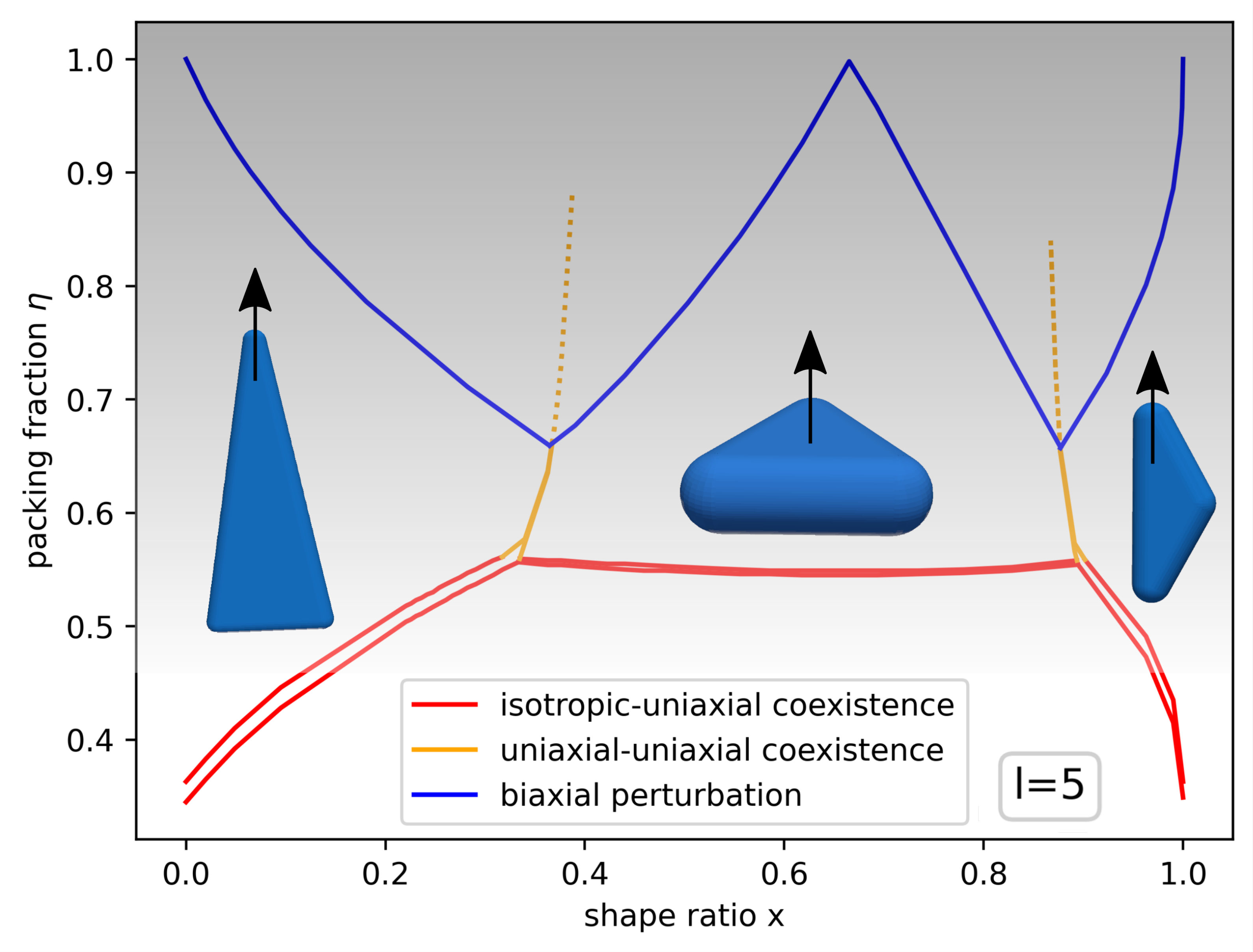}\hfill
    \includegraphics[width=0.485\textwidth]{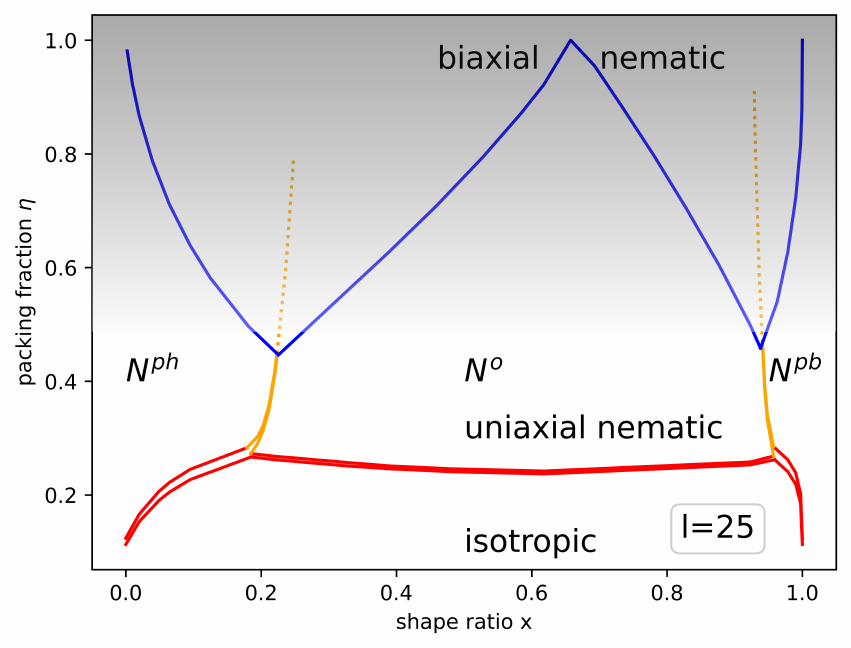}
    \caption{Phase diagram depicting the spatially homogeneous phases of hard spherotriangles at aspect ratios $l=5$ (left) and $l=25$ (right) in dependence of the packing fraction $\eta$ and the shape ratio $x$. The red lines indicate isotropic--uniaxial coexistence, the blue lines indicate uniaxial--biaxial coexistence and the orange lines indicate uniaxial--uniaxial coexistence. The latter becomes metastable within the biaxial regime and is thus shown as dotted lines to give an idea of what might be the dominant particle axis in the biaxial phase. As illustrated in Fig.~\ref{fig_coordinates}a, there are three stable orientations of the nematic director, which are visualized in the left diagram and marked with N$^\text{ph}$, N$^\text{o}$ and N$^\text{pb}$ in the right one.
    Once again, we shaded the area with $\eta\gtrsim0.45$,  where the used functional predicts stable phases with positional order in the limit of hard spherocylinders ($x=0$ or $x=1$), in light gray.}
    \label{fig_PD}
\end{figure*}

The homogeneous phase diagrams of hard cones and cylinders are shown in Fig.~\ref{fig_conecyl}, depending on the packing fraction $\eta$ and the rescaled aspect ratio $l/(1+l)$.
This rescaling of $l$ allows us to display all relevant information within a range from zero to one.
In the phase diagram,
 the dotted lines represent the isotropic-uniaxial coexistence densities for cylinders and the solid lines represent the coexistence densities for cones.
Both results are qualitatively similar.
For small (large) aspect ratios $l\ll1$ ($l\gg1$) the nematic phase becomes more stable upon further decreasing (increasing) $l$, such that the shape becomes more and more oblate (prolate).
For the more isotropic shapes at $l\approx1$ in between, no nematic order is possible and
the predicted transition region exceeds densities of presumed crystallization and even close packing, with an unphysical maximum at $\eta=1$ in the extreme case.
Moreover, both transitions are found to be of first order with comparable differences $\Delta\eta_\text{IN}$ of the packing fractions at coexistence, e.g., we find $\Delta\eta_\text{IN}\approx 0.001$ for cones and $\Delta\eta_\text{IN}\approx 0.002$ for cylinders with the same aspect ratio $l=0.1$.
As further expected, the phase behavior becomes asymptotically equal in the limits $l\rightarrow0$ and $l\rightarrow\infty$, where both shapes reduce to hard thin disks and Onsager rods \cite{OnsagerEffectsOfShape}, respectively, as studied with the present functional in Ref.~\onlinecite{wittmann2014}.

However, there are quantitative differences of the phase transition for the two shapes.
For example, the transition densities of the cylinder fluid peak at the unphysical value $\eta=1$ of the packing fraction for an aspect ratio of $l^*=1$, while this happens for $l^*\approx1.4$ in the case of cones.
This behavior is a direct mathematical consequence of the particle geometry: we can identify a \textit{most isotropic} shape (characterized by $l=l^*$) by noticing that the order parameter $S$ drops out of the chosen functional when putting together all tensorial components, such that no ordered phase can be described at all, compare appendices~\ref{app_fullWDhcone} and~\ref{app_fullWDhcyl}.
We are more interested in the regions, where the transition densities are lower and can thus be expected to describe the actual physical behavior.
For comparison, the functional used here predicts the onset of smectic order in a fluid of hard spherocylinders at packing fractions $\eta\gtrsim0.45$~\cite{wittmann2014} and we indicate this lower bound by using a gray background in Fig.~\ref{fig_conecyl}.
For all relevant aspect ratios, we predict that
the isotropic--nematic transition occurs at a larger packing fraction for cylinders than for cones.
As the volume of a cone is only one third of that of a cylinder, the particle number at a given  $\eta$ is larger, which results in  a stronger drive towards orientational order (in turn, the transition occurs at a lower particle number for cylinders).
This behavior is consistent with that shown in Fig.~2 of Ref.~\onlinecite{kubala2023splay} for hard connected spheres with different radii (mind that the aspect ratios of the bodies compared in this figure are not equal).

\subsection{Biaxial bodies: different directors \label{sec_RESbi}}

The homogeneous phase diagram of hard isosceles spherotriangles in Fig.~\ref{fig_PD} depicts the different transition densities as a function of the shape ratio $x$ for two aspect ratios $l=5$ (left) and $l=25$ (right).
We find that three distinct uniaxial phases, denoted by N$^\text{ph}$, N$^\text{o}$ and N$^\text{pb}$, which we model independently by choosing the main axis in the molecular frame to be
the triangle height, the face normal and the triangle base, respectively, are stable,  compare Fig.~\ref{fig_coordinates}.

The transition from the isotropic phase to a (uniaxial) nematic is always of first order.
In the two limiting cases $x=0$ and $x=1$, we recover the known result for hard spherocylinders,
which have been demonstrated to agree well with simulation results~\cite{wittmann2014}.
It is thus reassuring for our predictions to remain meaningful at intermediate values of $x$.
For increasing biaxiality of the particles (larger shape ratio when forming a N$^\text{ph}$ or smaller shape ratio when forming N$^\text{pb}$)
the packing fraction at which the transition occurs increases,
as the main axis becomes shorter, thus destabilizing (prolate) nematic order.
When the particle shape is sufficiently oblate, the stability limit of the isotropic phase is found at higher packing fractions than in the more prolate case and we eventually predict a transition to the N$^\text{o}$ phase.
Here, the corresponding transition densities are largely independent of the shape ratio.
At higher densities, we accordingly find two first-order
 transitions between uniaxial phases when increasing the shape ratio: first from N$^\text{ph}$ to N$^\text{o}$ and then from N$^\text{o}$ to N$^\text{pb}$.
In both cases, the transition region is slightly bent, such that the oblate phase destabilizes upon increasing the density.
Hence, there is a small range of shape ratios at which we observe two uniaxial phases following the
sequence isotropic to uniaxial oblate nematic to uniaxial prolate nematic and, finally, to biaxial nematic.

The biaxial phase, which we identify as a perturbation to simple uniaxial order (compare Sec.~\ref{sec_perturbedUAorder}), is most stable for shape ratios close to a uniaxial--uniaxial transition, which points to an equal weight of two distinct directors.
Moreover, the metastable uniaxial--uniaxial transition within the biaxial region (continued dotted lines) approximately indicates the dominant axis in the case of biaxial order (which we did not determine explicitly).
Our calculations are based on the assumption that the uniaxial--biaxial transition is of second order while we cannot fully rule out the possibility of a first-order transition within our current analytic treatment.
It is, however, quite reassuring that the biaxial perturbations of two distinct uniaxial phases at the uniaxial--uniaxial transition agree closely, suggesting that our calculations are consistent.
In contrast, we discuss in appendix~\ref{app_PA} that the more simplistic method to assume that one axis is perfectly ordered (compare Sec.~\ref{sec_perfectUAorder}) is not consistent for the different director orientations:
the stability of the biaxial phase is overestimated for taking N$^\text{ph}$ or N$^\text{pb}$ as the reference state but underestimated for N$^\text{o}$.
On the other hand, these more approximate transition lines (not included in Fig.~\ref{fig_PD}, but detailed in Fig.~\ref{fig_UB} below) are fully analytic and the second-order nature can explicitly be verified.

The qualitative behavior described above does not differ strongly between the two aspect ratios $l=5$ and  $l=25$ considered.
However, we find that the range of shape ratios $x$ which give rise to a stable N$^\text{o}$ phase increases due to the generally more oblate shape of spherotriangles with larger aspect ratio $l$.
More strikingly, the onset of biaxiality occurs at lower densities for these thinner particles,
as the shape also becomes less isotropic (in the opposite limit, $l=0$, a spherotriangle reduces to a perfect sphere).
This observation is quite important when contemplating the global stability of the biaxial phase, as we must take into account that other phases with positional order will probably preempt our predicted transitions at packing fractions $\eta\gtrsim0.45$ (as indicated by the fading backgrounds in Fig.~\ref{fig_PD}).
Hence, our results suggest that biaxial order should become stable for larger aspect ratios and, therefore, shape ratios closer to zero or one (for which oblate and prolate uniaxial phases coexist).
This conclusion is consistent with previous observations of biaxial nematic order in a system of extremely anisotropic biaxial particles \cite{dussi2018hard}.

\section{Conclusions}

In this work we have provided and applied a general recipe to investigate the homogeneous phase behavior of biaxial hard particle fluids within fundamental measure theory (FMT).
This framework allows us to determine all phase boundaries by solving independent algebraic equations.
In addition, we have demonstrated that treating biaxiality as a perturbation to uniaxial order constitutes a much more consistent procedure to determine the onset of the biaxial phase than the simple assumption of perfect uniaxial order.
Specifically, we have considered four order parameters that are established measures for orientational order in systems of biaxial particles.
While, in principle, even more order parameters will become important for more general shapes,
we found here that, upon choosing appropriate coordinate frames, all transitions between homogeneous phases can be consistently identified when taking into account only two parameters: the standard uniaxial order parameter $S$ and the order parameter $F$, which measures biaxiality.
We have investigated different particle shapes with a  relatively low symmetry, which still allowed for a detailed characterization of the phase behavior using the order parameters at hand.
As exemplified here for hard cones, uniaxial polar shapes typically require only one order parameter $S$ to describe bulk nematic order if the chosen coordinate systems are properly aligned with the symmetry axis.
The formation of a phase with global polar order can be most likely ruled out in our hard-core system for entropic reasons.
For biaxial shapes with $D_{2\mathrm{h}}$ symmetry (three mutually orthogonal symmetry planes) the four order parameters considered here are a standard choice \cite{rosso2007orientational}.
We have demonstrated here that no additional order parameters are needed to describe the biaxial nematic phase as long as all relevant axes are polar, which is the case for hard isosceles spherotriangles, but not for general hard spherotriangles.

Biaxial order can also emerge in mixtures of uniaxial bodies, specifically those involving both prolate and oblate species \cite{stroobants1984liquid,camp1996hard,do2010statistical}.
From the point of view of FMT, which naturally applies to mixtures without any conceptual complication, we expect that the phase biaxiality order parameter $P$ will play an important role in such a scenario.
A detailed investigation would be an interesting perspective for future work.
In turn, for arbitrary (convex) hard particles, FMT offers a straightforward way of identifying additional relevant order parameters through averages of products of distinct components of the rotation matrix~\eqref{eq_R} between the body frame and the lab frame, following the recipe outlined in Sec.~\ref{sec_wdrecipe}.
A comprehensive investigation of possible polar order using FMT would potentially require adding tensorial \cite{wittmann2014} or mixed \cite{wittmann2015FMMT} weighted densities to the current functional, or generalizing its expansion into spherical harmonics \cite{marechal2017density}, and performing a numerical minimization.

Although our present investigation is restricted to spatially homogeneous systems,
 FMT can also be applied to inhomogeneous situations,
 which are encountered in the presence of external walls~\cite{hansen2009edFMT,marechal2013density,schonhofer2018density},
 for free interfaces between coexisting homogeneous phases \cite{wittmann2014surface}
 or when more complex liquid crystal phases with positional order emerge \cite{wittmann2014,marechal2017density}.
In view of polar shapes, it would further be interesting to investigate the possibility of local polar order in adjacent splay domains~\cite{mertelj2018splay,sebastian2020ferroelectric,kubala2023splay} or
twist-bend or splay-bend structures~\cite{chaturvedi2019mechanisms,chiappini2021generalized,kotni2022splay}, scenarios in which the global orientational order remains nematic.
It is also worthwhile to calculate the Frank elastic coefficients~\cite{wittmann2015,de2016density} for different particles and investigate the effect of shape polarity and biaxiality on the elastic behavior.
An important issue concerns the global stability of the biaxial order predicted here: in Ref.~\onlinecite{cuetos2017phase} the transition to a biaxial nematic was found to be preempted by the onset of a smectic phase.
While we expect that the same happens for spherotriangles with $l=5$,  flatter particles, like those with $l=25$, were predicted here to exhibit biaxial behavior at lower packing fractions, such that it is more likely that this phase is actually stable for moderate densities (in particular for even larger values of $l$).
An elegant possibility to increase the stability range of the spatially homogeneous phases is to introduce slight modifications to the system, such as polydispersity \cite{belli2011polydispersity,chiappini2019} or depletion interactions~\cite{belli2012depletion}, which can destabilize positional order.
Also considering rounder shapes, which enhance the chance of a particle to slide out of a smectic layer, can favor the stability of a biaxial phase \cite{chiappini2019}.
Hence, one may suspect that a similar mechanism could be at work for smectics formed by spherotriangles (which require the particles to align in layers with an alternating up-down configuration), such that these could be less stable than those formed by spheroplatelets, which would be worthwhile to investigate in future work.

As a next step, the Brownian dynamics of the hard particles considered here could be explored by employing the present functional in a dynamical DFT (DDFT)~\cite{marconi1999DDFT,archer2004DDFT,teVrugtLW2020,teVrugtW2022}.
For example, in orientationally ordered phases, there is an anisotropic long-time diffusion which has been explored for uniaxial particles \cite{lowen1999anisotropic}, but not yet for biaxial particles.
As a computationally cheaper alternative for spatially inhomogeneous problems, phase field crystal (PFC) models \cite{ElderKHG2002,EmmerichWGTTG2012} are commonly used to study the dynamics in complex systems.
Here, the numerical effort is reduced by considering the dynamics of an orientation-averaged density field and different orientational order parameters rather than that of the full orientation-resolved density~\cite{WittkowskiLB2010,EmmerichWGTTG2012,teVrugtHKWT2022,teVrugtJW2021}.
Consequently, the results obtained here, which allow to express DFT functionals as a function of orientational order parameters in the biaxial case, are an excellent starting point to derive a PFC model for biaxial particles.
Finally, a further extension would be to investigate active biaxial particles \cite{wittkowski2012self},
which, also owing to the less symmetric particle shape, may exhibit circle swimming behavior \cite{KummeltHWBEVLB2013}.
Being a nonequilibrium system, this would again require the use of a dynamical theory.
By combining an appropriate FMT functional with the DDFT for biaxial active particles developed in Ref.\ \onlinecite{WittkowskiL2011}, the strategy presented in this work could be generalized to investigate the dynamics of the biaxial order parameters in the active system.

\section*{Acknowledgments}
The authors would like to thank Michael A.\ Klatt and Paul A.\ Monderkamp for helpful suggestions.
Funding by the Deutsche Forschungsgemeinschaft (DFG) under Project-ID 525063330 (MtV) and through the SPP 2265 under grant numbers LO 418/25-1 (HL) and WI 5527/1-1 (RW) is gratefully acknowledged.
AEM wants to thank the Studienstiftung des deutschen Volkes for financial support.

\appendix

\section{Calculation of weighted densities
\label{app_calculationWD}}

In this appendix, we provide additional details required to follow the general procedure, outlined in Sec.~\ref{sec_wdrecipe}, to calculate the weighted densities of the hard bodies shown in Fig.~\ref{fig_bodies}.

\subsection{Body parts \label{app_parameterization}}

We begin by presenting an appropriate parameterization for all body parts of hard cones, cylinders and spherotriangles, required as the second step in Sec.~\ref{sec_wdrecipe}.
This allows us to replace the integral $\int_{\partial\mathcal{B}}\upd\bvec{r}$ over the surface $\partial\mathcal{B}$ of the body in Eqs.~\eqref{eq_wdSCALAR}-\eqref{eq_wdTENSOR} by a sum of integrals corresponding to all contributing body parts, with respect to two parameters each, as specified below.
Here, we choose the parameterization, such that the symmetry axis of the cone and the cylinder points in the $z$-direction (i.e., it is parallel to $\vec{m}_3$). For the isosceles spherotriangles, we choose the height of the triangle, compare the left picture in Fig.~\ref{fig_coordinates}a.

\subsubsection{Cone mantle \label{app_cmp}}

We parameterize the surface of the cone's mantle in the form
\begin{equation}
\bvec{r}(t, \varphi)= \left(\begin{array}{c}  R(1-\frac{t}{L})\cos(\varphi)\\ R(1-\frac{t}{L})\sin(\varphi) \\ \frac{Ht}{L} \end{array}\right),
\label{eq_para_CONEMANTLE}
\end{equation}
where $L$ is defined as
\begin{equation}
    L:=\sqrt{H^2+R^2}.
\end{equation}
Here, $R=D/2$ is the radius of the cone and $H$ is its height. We use $\varphi$ and $t$ to parameterize the cone surface, where $\varphi$ is the polar angle and $t$ gives the distance along the cone mantle from the base of the cone. Therefore, $t \in [0, L]$ and $\varphi \in [0, 2 \pi]$.

\subsubsection{Disk}
The surface of a disk of radius $R=D/2$ is parameterized by polar coordinates given by two parameters $r\in [0, R]$ and $\varphi \in [0, 2\pi]$. Therefore,
\begin{equation}
    \bvec{r}(r, \varphi)=\left(\begin{array}{c}
         r \cos\left(\varphi\right)\\
         r \sin\left(\varphi\right)\\
         0
    \end{array}\right).
\end{equation}
The cone only requires one disk with the normal vector pointing down, while the cylinder requires two disks with same parameterizations but opposite normal vectors.

\subsubsection{Circular ring (torus)}
A small complication arises for
the cone and the cylinder, as these bodies contain points with infinite curvature, where the normal vector is not well defined.
To resolve this issue at the circular ring between mantle and capping disks, we consider appropriate parts of a torus with width $r$ and radius $R=D/2$.
Then, after defining all necessary geometric quantities, we take the limit $r \to 0$.
For the calculation of the weighted densities, we use standard torus coordinates, which are defined as
\begin{equation}
\bvec{r}\left(\varphi, \xi\right)=\left(
\begin{array}{c}
(R+r\cos(\xi) )\cos(\varphi) \\
(R+r\cos(\xi) )\sin(\varphi) \\
r\sin(\xi)
\end{array}
\right)
\end{equation}
where $\varphi \in [0, 2 \pi]$ and, in the case of the cone, $\xi \in [0, \arccos\left(-R/L\right)]$.
For hard cylinders, we need two tori with $\xi \in [0, \pi/2]$ and $\xi \in [\pi/2, \pi]$.
When the corresponding weighted densities are calculated (according to the six steps in Sec.~\ref{sec_wdrecipe}), we finally set $r \to 0$.

\subsubsection{Sphere}
The surface of a sphere can be parameterized by using the spherical coordinates
\begin{equation}
\bvec{r}\left( \vartheta, \varphi\right)=
\left(\begin{array}{c}
    R\sin( \vartheta) \cos( \varphi) \\
    R\sin( \vartheta) \sin( \varphi) \\
    R\cos( \vartheta)
\end{array}\right)
\end{equation}
and fixing the radial coordinate $R$.
For the cone, we require a part of a sphere in the limit of $R \to 0$ on its tip, which allows us to calculate well-defined geometrical quantities, as explained above for the circular ring.
In this case, we have $\vartheta \in [0, \pi-\arccos\left(-R/L\right)]$  and $\varphi \in [0, 2 \pi]$.
In practice, a sphere in the limit $R \to 0$ only contributes to the integrated Gaussian curvature and thus only to the scalar weighted density $n_0$ (together with the circular ring), which can be set directly to $n_0=\rho$, as the one is a simply connected body.
For spherotriangles, there are three parts of a sphere with radius $R=D/2$ on their edges.
As these parts will always add up to a full sphere, we can simply assume $\varphi \in [0, 2 \pi]$ and $\vartheta \in [0, \pi]$ for all contributions at once.

\subsubsection{Cylinder Mantle}
In our calculations for a hard cylinder, we parameterize  the cylinder mantle by choosing a parameter $h\in[0, H]$, where $H$ is the full height of the cylinder and $\varphi \in[0, 2\pi]$ is an angle. This equivalent to using standard cylinder coordinates with a fixed radial coordinate $R=D/2$. This corresponds to the parameterization
\begin{align}
\bvec{r}\left(h, \varphi\right)=\left(\begin{array}{c}
    R \cos \left(\varphi\right) \cr
    R \sin \left(\varphi\right) \cr
    h
\end{array}\right)
\end{align}
For spherotriangles, we must consider three parts of cylinders with $h \in [0,A]$ or $h \in [0,B]$ and always $\varphi \in [-\pi/2, \pi/2]$.
Moreover, for each part, $\bvec{r}\left(h, \varphi\right)$ needs to be rotated in the $xz$-plane.
This rotation angle is $-\pi/2$ for the base line of length $A$,
and $\gamma$ or $\pi-\gamma$ for the other two sides of length $B$.

\subsubsection{Triangle}

Flat isosceles triangles are parameterized with Cartesian coordinates, therefore
\begin{equation}
    \bvec{r}\left(x, z\right)=\left(\begin{array}{c}
         x \\
         0 \\
         z
    \end{array}\right),
\end{equation}
where $x \in [-A/2, A/2]$ and $z \in [0, \sqrt{B^2-A^2/4}(1-2|x|/A)]$. We require in total two triangles with opposite normal vectors.

\subsection{Geometrical measures \label{app_measurescone}}

Using the different parameterizations of the relevant body parts,
we can calculate all geometric quantities in Eqs.~\eqref{eq_wdSCALAR}-\eqref{eq_wdTENSOR}, i.e., the two principal curvature directions $\bvec{v}_1$ and $\bvec{v}_2$, the surface unit normal vector $\bvec{n}$, the two principal curvatures $\kappa_1$  and $\kappa_2$, the Gaussian curvature $\mathcal{G}$ and the mean curvature $\mathcal{H}$.
This third step in Sec.~\ref{sec_wdrecipe}, is carried out below explicitly for the exemplary case of a cone mantle, whose surface is parameterized in Eq.~\eqref{eq_para_CONEMANTLE}.

\subsubsection{Unit vectors}

To calculate the unit vectors, we take the derivative of the vector $\bvec{r}(t, \varphi)$, given in Eq.~\eqref{eq_para_CONEMANTLE}, with respect to $t$ and $\varphi$, which yields
   \begin{equation}
      \frac{\partial\bvec{r}}{\partial \varphi}=\left(\begin{array}{c}
       -R(1-\frac{t}{L}) \sin(\varphi)\\ R(1-\frac{t}{L}) \cos(\varphi) \\ 0
      \end{array}\right)
  \end{equation}
  and
  \begin{equation}
      \frac{\partial\bvec{r}}{\partial t}=\left(\begin{array}{c}
       -\frac{R}{L}\cos(\varphi)\\ -\frac{R}{L}\sin(\varphi) \\ \frac{H}{L}
      \end{array}\right).
      \label{eq_drdt}
  \end{equation}
Both of these two vectors are tangential to the cone's surface and perpendicular to each other.
Therefore, these vectors, after being normalized, yield the expressions
     \begin{equation}
      \bvec{v}_{1}=\left(\begin{array}{c}
       -\sin(\varphi)\\ \cos(\varphi) \\ 0
      \end{array}\right).
  \end{equation}
  and
   \begin{equation}
   \bvec{v}_{2}=\left(\begin{array}{c}
       -\frac{R}{L}\cos(\varphi)\\ -\frac{R}{L}\sin(\varphi) \\ \frac{H}{L}
      \end{array}\right),
  \end{equation}
and are precisely the vectors $\bvec{v}_{1}$ and $\bvec{v}_{2}$
we have been looking for.

Going on, we find that
\begin{equation}
    \frac{\partial \bvec{r}}{\partial \varphi}\times\frac{\partial \bvec{r}}{\partial t}=\frac{R}{L}\left( \begin{array}{c}
       H\left(1-\frac{t}{L}\right)\cos(\varphi)    \\ H\left(1-\frac{t}{L}\right)\sin(\varphi) \\ R \left(1-\frac{t}{L}\right)
       \end{array}\right).
\end{equation}
This vector is necessarily perpendicular to the cone's surface.
After normalizing it, we get our required normal vector, which is given by
\begin{equation}
    \bvec{n}=\left(\begin{array}{c}\frac{H}{L}\cos(\varphi)\\ \frac{H}{L}\sin(\varphi) \\ \frac{R}{L}\end{array}\right).
\end{equation}

\subsubsection{Weingarten map and curvatures}
Next, we determine the components of the Weingarten map, which allows us to calculate the principal curvatures of the cone.
In order to do so, the first step is to calculate the metric tensor $g$ of the cone's mantle.
The $ij$th component of the metric tensor is defined as
\begin{equation}
    g_{ij}:=\frac{\partial \bvec{r}}{\partial x_{i}} \cdot \frac{\partial \bvec{r}}{\partial x_{j}}\,,
    \label{eq_gij}
\end{equation}
where $x_{i},x_{j}\in\{\varphi,t\}$.
Following this definition,
we arrive at
\begin{equation}
    g_{11}=\frac{\partial \bvec{r}}{\partial t}\cdot\frac{\partial \bvec{r}}{\partial t} = \frac{R^2}{L^2}\left(\cos(\varphi)^2+\sin(\varphi)^2\right)+\frac{H^2}{L^2}=1
\end{equation}
and
\begin{align}
    g_{22}=\frac{\partial \bvec{r}}{\partial \varphi}\cdot\frac{\partial \bvec{r}}{\partial \varphi} &= R^2\left(1-\frac{t}{L}\right)^2\left(\cos(\varphi)^2+\sin(\varphi)^2\right) \!\!\!\!\!\!\!\!\cr
    &=R^2\left(1-\frac{t}{L}\right)^2,
\end{align}
while the cross terms
\begin{equation}
    g_{12}=g_{21}=\frac{\partial \bvec{r}}{\partial t}\cdot\frac{\partial \bvec{r}}{\partial \varphi}=0
\end{equation}
 vanish.
Therefore, the metric tensor of the cone's mantle can be represented by the matrix
\begin{equation}
g=\left( \begin{array}{rrr}
1 & 0\\
0 & R^2\left(1-\frac{t}{L}\right)^2\\
 \end{array}\right),
\end{equation}
  whose inverse is given by
\begin{equation}
g^{-1}=\left( \begin{array}{rrr}
1 & 0\\
0 & \frac{1}{R^2\left(1-\frac{t}{L}\right)^2}\\
 \end{array}\right).
  \label{eq_G}
\end{equation}

In order to calculate the components of the Weingarten map, we need a second matrix. Its elements are defined by
\begin{equation}
    B_{ij}=\frac{\partial \bvec{n}}{\partial x_{i}} \cdot \frac{\partial \bvec{r}}{\partial x_{j}},
\end{equation}
where $x_{i},x_{j}\in\{\varphi,t\}$. We calculate
\begin{equation}
    \frac{\partial \bvec{n}}{\partial \varphi}=\left(\begin{array}{c}-\frac{H}{L}\sin(\varphi) \\ \frac{H}{L}\cos(\varphi) \\ 0 \end{array}\right)\,,\ \ \
    \frac{\partial \vec{n}}{\partial t}=\vec{0}\,.
\end{equation}
As we have calculated all required derivatives, we are now ready to calculate the whole matrix.
Both $B_{11}=0$ and $B_{12}=0$ vanish immediately, since they contain the vanishing factor $\partial \vec{n}/\partial t$.
Moreover, $B_{21}=0$ vanishes since $\partial \vec{n}/\partial \phi$ is perpendicular to $\partial \vec{t}/\partial t$ from Eq.~\eqref{eq_drdt}.
Thus, only
\begin{equation}
   B_{22} = \frac{\partial \bvec{n}}{\partial \varphi}\cdot \frac{\partial \bvec{r}}{\partial \varphi}=\frac{H}{L}R\left(1-\frac{t}{L}\right)\,
\end{equation}
 contributes a nontrivial result, such that
\begin{equation}
B=\left( \begin{array}{rrr}
0 & 0\\
0 & \frac{H}{L}R\left(1-\frac{t}{L}\right)\\
 \end{array}\right).
 \label{eq_B}
\end{equation}
Finally, the matrix representation of the Weingarten map is given by
\begin{equation}
    W = g^{-1}\cdot B =
\left( \begin{array}{rrr}
0 & 0\\
0 & \frac{H}{LR\left(1-\frac{t}{L}\right)}\\
\end{array}\right),
\end{equation}
where we have inserted Eqs.~\eqref{eq_G} and~\eqref{eq_B}.

From differential geometry, it is known that the principal curvatures are just the eigenvalues of the Weingarten map.
Since $W$ is a diagonal matrix, the eigenvalues are $\kappa_{2}=0$ with principal direction $\bvec{v}_{2}$
and  $\kappa_{1}=\frac{H}{LR\left(1-t/L\right)}$ with principal direction $\bvec{v}_{1}$.
Therefore, the cone mantle has the Gaussian curvature
\begin{equation}
    \mathcal{K}=\kappa_{1} \kappa_{2}=0
\end{equation}
and the mean curvature
\begin{equation}
    \mathcal{H}=\frac{1}{2}(\kappa_{1}+\kappa_{2})=\frac{1}{2}\frac{H}{LR\left(1-\frac{t}{L}\right)}\,,
\end{equation}
which depends on the position $t$ on the cone's mantle.
As expected, the  curvatures are identical to those of a cylinder in the limit $H \to \infty$, $L \to \infty$ with $H/L \to 1$ and finite $R$.

\section{Weighted densities \label{app_fullWD}}

 \subsection{Hard cones \label{app_fullWDhcone}}
We now present the full set of relevant weighted densities for hard cones with height $H$ and base diameter $D=2R$.
For simplicity, we neglect the order parameter $P$ by assuming $P=0$, as it is irrelevant for the homogeneous bulk phase behavior for our choice of coordinates (compare the discussion in Sec.~\ref{sec_cosy}).
The contribution of $P$ and other order parameters will be explicitly shown in appendix~\ref{app_fullWDhsc} for hard spherocylinders.

We begin by giving the scalar weighted densities
\begin{align}
n_{0}&=\rho\,, \cr
n_{1}&=\frac{\rho}{4}\left(H+R\,\arccos\!\left(\frac{-R}{\sqrt{H^2+R^2}}\right)\right),\cr
n_{2}&=\rho\, \pi R(R+\sqrt{H^2+R^2})\,,\cr
n_{3}&=\frac{\rho}{3}\pi R^2 H\,.
\end{align}
For our choice of coordinates, only the third component
\begin{align}
(\overrightarrow{n}_1)_{3}&=\rho\,R\,\frac{H-R-\sqrt{H^2+R^2}}{4\sqrt{H^2+R^2}} \,\langle\cos(\theta)\rangle
\end{align}
of the first vector weighted density is nonzero, where $\langle\cos(\theta)\rangle=\int \upd\mathbf{O}  \,g(\phi, \theta,\psi)\,\cos(\theta)$ denotes the orientational average of $\cos(\theta)$, as in Eq.~\eqref{eq_orderparameters}, which is zero for (apolar) nematic order.
Since $\overrightarrow{n}_{2}=\overrightarrow{0}$, the vectors do not contribute overall, such that there is no order parameter like $\langle\cos(\theta)\rangle$ that is sensitive to polar order in the functional employed here.
In view of other shapes, we learn here that the vectorial contributions do not necessarily support  polar order  even if the particle shape is polar,
such that it appears to be a safe approximation to generally neglect these terms in our study.

The diagonal tensor weighted densities read
\begin{align}
(\overleftrightarrow{n}_1)_{11} &= (\overleftrightarrow{n}_1)_{22}\cr
&=\frac{\rho}{24} \left(H\, \frac{5H^2-2R^2}{\left(H^2 + R^2\right)} - 3 R \hat{x}\right) S ,\cr
(\overleftrightarrow{n}_1)_{33} &= \frac{\rho}{24} \left(H\, \frac{4R^2-10H^2}{\left(H^2 + R^2\right)} + 6 R \hat{x}\right)S,
\end{align}
where
\begin{equation}
    \hat{x}:=\frac{1}{2}\left(\arccos\!\left(-\frac{R}{\sqrt{H^2+R^2}}\right)-\frac{HR}{H^2+R^2}\right)
\end{equation}
and
\begin{align}
(\overleftrightarrow{n}_2)_{11} &=(\overleftrightarrow{n}_2)_{22}\cr
&= \frac{\rho}{6} \left( \left( -2\pi R^2 + \pi R \frac{H^2-2R^2}{\sqrt{H^2 + R^2}}  \right)S\right. \cr
&\quad\quad\quad\  +  2 \pi R^2 + 2\pi R\sqrt{H^2+R^2}  \bigg)\,,\cr
(\overleftrightarrow{n}_2)_{33} &= \frac{\rho}{3}\left(  \left( 2 \pi R^2 - \pi R \frac{H^2-2R^2}{\sqrt{H^2 + R^2}}  \right)S\right. \cr
&\quad\quad\quad\ \  +    \pi R^2 + \pi R \sqrt{H^2+R^2}  \bigg)\,,
\end{align}
while all nondiagonal elements vanish.

\subsection{Hard cylinders \label{app_fullWDhcyl}}
Here, we give the full set of weighted densities for hard cylinders~\cite{wittmann2014} with height $H$ and diameter $D$, setting again $P=0$.
We start by giving the scalar weighted densities
\begin{align}
n_{0} &= \rho\,, \cr
n_{1} &= \frac{\rho}{8}\left(2H+\pi D\right), \cr
n_{2} &= \frac{\rho}{2}\left(2\pi H D+\pi D^2\right), \cr
n_{3} &= \frac{\rho}{4}\,\pi H D^2 \,,
\end{align}
while the vectors vanish for this apolar shape.
The diagonal components of the  $(\overleftrightarrow{n}_1)$ tensor are
\begin{align}
(\overleftrightarrow{n}_1)_{11} &=(\overleftrightarrow{n}_1)_{22}= -\frac{1}{2}(\overleftrightarrow{n}_1)_{33}\cr
 &=\frac{\rho}{32}\left(4H-\pi D\right)S.
\end{align}
The diagonal components of the $(\overleftrightarrow{n}_2)$ tensor are
\begin{align}
(\overleftrightarrow{n}_2)_{11} &= (\overleftrightarrow{n}_2)_{22} \cr
&= \frac{\rho}{6}\left(\pi HD(2+S)+\pi D^2(1-S)\right), \cr
(\overleftrightarrow{n}_2)_{33} &= \frac{\rho}{6}\left(2\pi HD(1-S)+\pi D^2(2S+1)\right).
\end{align}
Again, all nondiagonal elements vanish.
The total contribution of the tensors to $\phi_2$ in Eq.~\eqref{eq_phi2zeta} is
\begin{align}\label{eq_wdcylinder}
 \mbox{Tr}\!\left[\overleftrightarrow{n}_1\overleftrightarrow{n}_2\right] = \rho^2\pi D\frac{(H-D)(4H-\pi D)}{32}S^2\,,
  \end{align}
which vanishes for $H=D$ (and $4H=\pi D$).
It can also be shown that the expression in Eq.~\eqref{eq_phi3TR} vanishes for $H=D$, such that the dependence on $S$ drops out of the functional for such a shape.

\subsection{Hard spherotriangles \label{app_fullWDhst}}

The scalar weighted densities of hard isosceles spherotriangles with diameter $D$, base length $A$ and two sides of length $B$ read
\begin{align}
n_{0} &= \rho, \cr
n_{1} &= \frac{\rho}{8} \left(A+2B+4D\right), \cr
n_{2} &= \frac{\rho}{2} \left(\pi D(A+2B)+2\pi D^2 + A \sqrt{4B^2-A^2}\right), \cr
n_{3} &= \frac{\rho}{24} \left(3\pi D^2(A+2B)+4\pi D^3+12D A \sqrt{4B^2-A^2}\right),\cr
\label{eq_WDHSTscalar}
\end{align}
where the contributions which are quadratic, linear and constant in $D$ originate from the spherical caps, parts of cylinder mantles and triangles, respectively.
In general, the vectors do yield nonzero contributions, which vanish for the apolar phases of interest (see the discussion in appendix~\ref{app_fullWDhcone}) and are omitted here due to their lengthy general form.
Only in the special case $A=B$ of equilateral spherotriangles, we have $\overrightarrow{n}_{2}=\overrightarrow{0}$ such that the vectors are truly irrelevant.

The tensor weighted densities explicitly depend on the shape ratio $x$ of the spherotriangles through  $\gamma$, which  is the half opening angle of the triangle, see Fig.~\ref{fig_bodies}.
The nonvanishing diagonal components of $\overleftrightarrow{n}_1$ have only contributions from the cylindrical parts and read
\begin{align}
(\overleftrightarrow{n}_1)_{11}  &=(\overleftrightarrow{n}_1)_{22}(S,U,-P,-F)
\cr&= \frac{\rho}{32} \left( A(-S+\sqrt{3}U+\sqrt{3}P-3F)\right.\cr
&\quad \ \ \ \ \    +2B \left( (-S+\sqrt{3}U+\sqrt{3}P-3F)\sin(\gamma)^2\right.\cr
&\ \ \ \ \ \ \ \ \ \ \ \ \ \ \ \ \ \left.\left.+2(S-\sqrt{3}P)\cos(\gamma)^2
 \right)\right)
\label{eq_n111str}
\end{align}
and
\begin{align}
(\overleftrightarrow{n}_1)_{33} &= \frac{\rho}{16} \left(A(S-\sqrt{3}U) \right.\cr
&\quad \ \ \ \ \  \left. +2B\left( (S-\sqrt{3}U)\sin\left(\gamma\right)^2
 - 2S \cos\left(\gamma\right)^2 \right)\right).\cr
\end{align}
The relation between $(\overleftrightarrow{n}_1)_{11}$ and $(\overleftrightarrow{n}_1)_{22}$ in Eq.~\eqref{eq_n111str} stems from the fact that these tensor components are related by a polar rotation of the body in the lab frame, achieved by a
redefinition $\phi\rightarrow\phi\pm\pi/2$ of the polar angle~$\phi$.
Just like $n_2$, the tensor $\overleftrightarrow{n}_2$ has contributions from all body parts.
Its relevant diagonal components read
\begin{align}
&(\overleftrightarrow{n}_2)_{11}
=  (\overleftrightarrow{n}_2)_{22}(S,U,-P,-F)\cr
&=\frac{\rho}{24} \left( \pi AD( -S+\sqrt{3}U+\sqrt{3}P-3F +4 )\right.\cr
&\quad\quad\quad + 2\pi BD \left(  (-S+\sqrt{3}U+\sqrt{3}P-3F)\sin(\gamma)^2   \right.\cr
&\quad\quad\quad\quad\quad\quad\quad\  +  \left.  2\,(S-\sqrt{3}P)\cos(\gamma)^2 +4\right)\cr
&\quad\quad\quad + \left. 8\pi D^2\right.\cr
&\quad\quad\quad + \left.  2A\sqrt{4B^2-A^2}(S+\sqrt{3}U-\sqrt{3}P-3F+2) \right)\cr
\end{align}
and 
\begin{align}
(\overleftrightarrow{n}_2)_{33}
&= \frac{\rho}{12} \left( \pi AD (2+S-\sqrt{3}U)\right.\cr
&\ \ \ \ \ \ \ \ \ + 2\pi BD \left((S-\sqrt{3}U)\sin(\gamma)^2\right.\cr
&\ \ \ \ \ \ \ \ \ \ \ \ \ \ \ \ \ \ \ \ \ \ \left.  -2S\cos(\gamma)^2 +2\right)  \cr
&\ \ \ \ \ \ \ \ \ \left.  + 4\pi D^2  \right.\cr
&\ \ \ \ \ \ \ \ \ \left. + 2A\sqrt{4B^2-A^2}(-S-\sqrt{3}U+1) \right).\cr
\label{eq_n233str}
\end{align}
To determine the tensorial weighted densities for other orientations of the primary director, we can use the substitutions from \cref{eq_substTAU1} or \cref{eq_substTAU2}, as detailed in Sec.~\ref{sec_cosy}.
For example, aligning the $z$-axis of the body frame with the base line of the triangle, compare the right picture in Fig.~\ref{fig_coordinates}b, we get the first tensor component
\begin{align}
(\overleftrightarrow{n}_1)_{11} &= \frac{\rho}{16} \left( A\,(S-\sqrt{3}P)\right.\cr
&\quad \ \ \ \ \    +B \left( 2(S-\sqrt{3}P)\sin(\gamma)^2
\right.\cr
&\ \ \ \ \ \ \ \ \ \ \ \ \ \ \ \left.\left.+ (-S+\sqrt{3}U+\sqrt{3}P-3F)\cos(\gamma)^2
 \right)\right) \cr
\label{eq_n111strX}
\end{align}
upon applying \cref{eq_substTAU2} to \cref{eq_n111str}.

\begin{figure*}[t]
    \centering
    \includegraphics[width=0.32\textwidth]{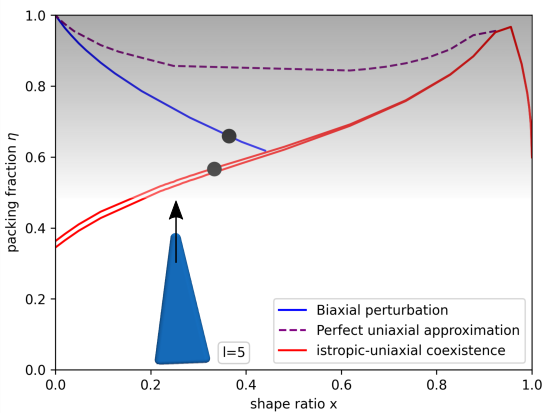}
    \includegraphics[width=0.32\textwidth]{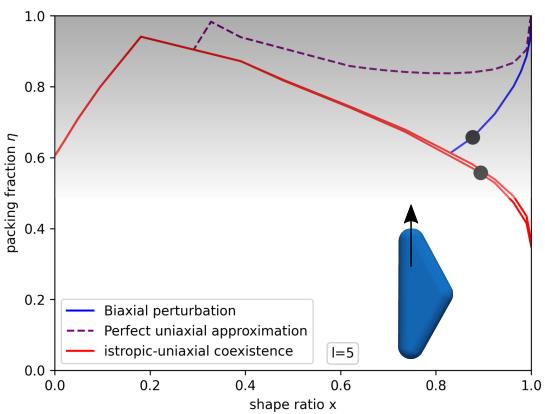}
    \includegraphics[width=0.32\textwidth]{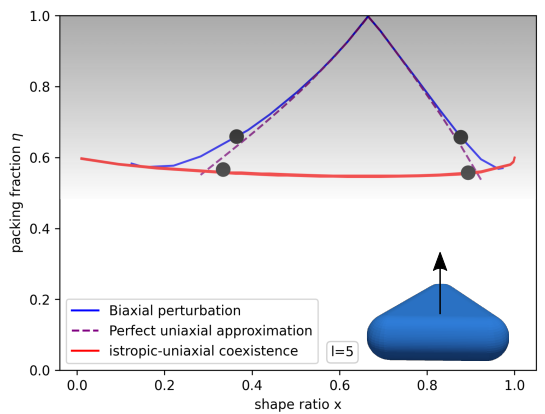}
    \caption{Phase diagrams of hard spherotriangles upon imposing the three different orientations of the uniaxial director illustrated in \cref{fig_coordinates}, as indicated by the drawn bodies.
    In each plot, the red lines indicate the isotropic--uniaxial coexistence densities calculated by free minimization, following Sec.~\ref{sec_uniaxial}, and the blue lines indicate the uniaxial--biaxial transition identified by a perturbation of simple uniaxial order as outlined in Sec.~\ref{sec_perturbedUAorder}.
    In addition to these results, which are also shown in Fig.~\ref{fig_PD},
    we compare the predictions of the perfect uniaxial approximation (dotted purple line) to locate the uniaxial--biaxial transition, as outlined in Sec.~\ref{sec_perfectUAorder}.
    The black dots mark the triple points, where two uniaxial phases with different director orientations coexist with either the isotropic or the biaxial phase, compare Fig.~\ref{fig_PD}.
    The phase diagrams presented here thus also include the regions beyond these points, where the given director orientation is unstable.
    }
    \label{fig_UB}
\end{figure*}

\subsection{Limits of hard spherocylinders \label{app_fullWDhsc}}

In the limits $x=0$ or $x=1$ of extreme shape ratios, a spherotriangle turns into a  spherocylinder.
Therefore, by evaluating the weighted densities of hard spherotriangles from appendix~\ref{app_fullWDhst} for these special cases,
we can directly obtain the full set of weighted densities of hard spherocylinders for two different choices of the body frame, which results in a different dependence on the order parameters $S$, $U$, $P$ and $F$, as we elaborate below.
Recall that we have used the convention, illustrated in the left picture in Fig.~\ref{fig_coordinates}a, that the height of the triangle points in the $z$-direction of the body frame.

In the first case, $x=0$, we recover the standard convention used for uniaxial bodies.
By setting $A=0$ and $L=B$ in Eqs.~\eqref{eq_n111str}-\eqref{eq_n233str}, we obtain the tensor components
\begin{align}
(\overleftrightarrow{n}_1)_{11} &= \rho  L \left( \frac{S}{8} - \frac{\sqrt{3}P}{8} \right), \cr
(\overleftrightarrow{n}_1)_{22} &= \rho  L \left( \frac{S}{8} + \frac{\sqrt{3}P}{8} \right), \cr
(\overleftrightarrow{n}_1)_{33} &= \rho  L \left( -\frac{S}{4} \right), \cr
(\overleftrightarrow{n}_2)_{11} &= \rho \left( \frac{D^2 \pi}{3} + \frac{D L \pi}{6} \left( 2 + S - \sqrt{3} P \right) \right), \cr
(\overleftrightarrow{n}_2)_{22} &= \rho \left( \frac{D^2 \pi}{3} + \frac{D L \pi}{6} \left( 2 + S + \sqrt{3} P \right) \right), \cr
(\overleftrightarrow{n}_2)_{33} &= \rho \left( \frac{D^2 \pi}{3} + \frac{D L \pi}{6} \left( 1 - S \right) \right)\,,
\label{eq_wdHSC_SP}
\end{align}
while the total contribution of the tensors to $\phi_2$ in Eq.~\eqref{eq_phi2zeta} is
\begin{align}\label{eq_wdscylinder}
 \mbox{Tr}\!\left[\overleftrightarrow{n}_1\overleftrightarrow{n}_2\right] = \rho^2\frac{D L^2 \pi}{8}\left(S^2+3 P^2\right).
  \end{align}
As we can see, within our general treatment, there is a dependence on $P$ in addition to the standard nematic order parameter $S$.
Setting $P=0$, these weighted densities reduce to the common expressions previously reported in the literature (we also recover the orientation-independent scalar weighted densities which are not repeated here) \cite{hansen2010tensorial,wittmann2014,wittmann2016,wittmann2015}.
Hence, neglecting $P$ has already been demonstrated to be well justified when one is only interested in the bulk phase behavior, which we have explicitly verified in Sec.~\ref{sec_RESop}.
The form of the weighted densities in Eq.~\eqref{eq_wdHSC_SP}
is, however, helpful to understand physical scenarios where the uniaxial symmetry of the phase is broken, for example due to external fields.
Most importantly, we anticipate the explicit importance of $P$ for the bulk phase behavior of (uniaxial) rod-disk mixtures.

While we have seen that $P$ contributes in general to the weighted densities of uniaxial bodies, $U$ and $F$ do not appear in Eq.~\eqref{eq_wdscylinder} because a spherocylinder is not a biaxial particle.
However, in the second case, $x=1$, the symmetry axis of the resulting spherocylinder does not coincide with the orientation in the body frame, which means that we formally treat the body as if it was biaxial.
Indeed, if we set $L$=$A$=$2B$ in Eqs.~\eqref{eq_n111str}-\eqref{eq_n233str}, we arrive at the following contribution of the tensorial weighted densities
\begin{align}\label{eq_wdscylinder2}
 \mbox{Tr}\!\left[\overleftrightarrow{n}_1\overleftrightarrow{n}_2\right] = \rho^2\frac{D^2 L \pi}{32}\left((\sqrt{3}U-S)^2+3(\sqrt{3}F-P)^2\right),
  \end{align}
  which depends on all four order parameters (the same result can be obtained from \cref{eq_wdscylinder} by performing the substitutions in \cref{eq_substTAU2}).
These weighted densities still contain the same physics, but one has to give up the interpretations of the order parameters outlined in Sec.~\ref{sec_cosy}.
In our particular example, this means that $S$ does not appropriately describe
the uniaxial order of hard spherocylinders when setting $U=P=F=0$ in \cref{eq_wdscylinder2}.

To show the difference between the two descriptions of hard spherocylinders, we show in Fig.~\ref{fig_UB} the predicted isotropic--uniaxial transition lines (only allowing for nonzero values of $S$) of hard spherotriangles for the full range of shape ratios $x$ of each chosen director orientation.
This compilation includes results which are unstable with respect to a different director orientation and thus not contained in Fig.~\ref{fig_PD}.
The left plot in Fig.~\ref{fig_UB} shows that the stable spherocylinder limit of isosceles spherotriangles aligned along the direction of their height is obtained for $x=0$ with Eq.~\eqref{eq_wdscylinder}, while the transition for $x=1$ with Eq.~\eqref{eq_wdscylinder2} is located at much higher densities and in the unstable regime.
The same unphysical result is found in both limits $x=0$ and $x=1$ for isosceles spherotriangles aligned along the normal direction to their face (central plot in Fig.~\ref{fig_UB}) and in the limit $x=0$ for isosceles spherotriangles aligned along their triangle base (right plot in Fig.~\ref{fig_UB}).
In the latter case, taking $x=1$ again yields the stable spherocylinder limit.

\subsection{Comments on general hard spherotriangles \label{app_gentriangle}}

For previous calculations, due to the symmetry of isosceles spherotriangles, all nondiagonal terms of the tensor weighted densities were automatically zero.
However, for arbitrary spherotriangles with side lengths $A$, $B$ and $C\neq B$, there may be both such nonzero cross terms and additional order parameters resulting from
orientational averages of mixed terms $\langle\hat{\mathcal{R}}_{ij}\hat{\mathcal{R}}_{kl}\rangle$ with $i\neq k$ and/or $j\neq l$ of the rotation matrix, which are not included in Eq.~\eqref{eq_Rii2av}, since these correspond to cross terms of the Saupe matrix.

To give an example of such cross terms, we consider the limit of perfect uniaxial order, as introduced in Sec.~\ref{sec_perfectUAorder}, for a general spherotriangle with the nematic director parallel to the face normal of the triangle, compare the central picture in Fig.~\ref{fig_coordinates}a.
Then, the nondiagonal elements of the tensorial weighted densities can be given in the compact form
\begin{align}
(\overleftrightarrow{n}_1)_{12} &=(\overleftrightarrow{n}_1)_{21}\cr
&= -\frac{3\rho}{16} \bigg(B \sin\left(\mu\right) \cos\left(\mu \right) S_{2d} \cr
&\quad \ \ \ \ \ \ \ \ \  +C \sin(\eta) \cos(\eta) S_{2d} \bigg) \ \ \ \ \ \
\end{align}
and
\begin{align}
(\overleftrightarrow{n}_2)_{12} &=(\overleftrightarrow{n}_2)_{21}\cr
&= -\frac{\rho}{4} \bigg( \pi D B \sin(\mu) \cos(\mu) S_{2d} \cr
&\quad \ \ \ \ \ \ \ \, + \pi D C \sin(\eta) \cos(\eta) S_{2d}\bigg),\ \ \ \ \ \ \
\end{align}
where $S_{2d}$, as defined in Eq.~\eqref{eq_orderparameter2d}, is identified according to Eq.~\eqref{eq_OPsPA} and the angles $\mu$ and $\eta$ are defined as
\begin{align}
     \mu&=\arccos\bigg(\frac{B^2+A^2-C^2}{2BA}\bigg),\cr
     \eta&=\arccos\bigg(\frac{B^2+C^2-A^2}{2BC}\bigg)+\mu.
\end{align}
Note that the cross terms presented here vanish for isosceles spherotriangles, i.e., if we set $B=C$.

\section{Details on the uniaxial--biaxial transition \label{app_PA}}

In the main text, we outline two different methods to determine the location of the uniaxial--biaxial transition.
First, in the perfect uniaxial approximation, we assume $S=1$ and $P=U=0$,
which leaves us with a single order parameter $F\simeq S_{2d}$ and we can analytically minimize the basic equations of DFT as for the functional in two spatial dimensions. We explain this in greater detail in Sec.~\ref{sec_perfectUAorder}.
We also use a perturbative approach, where we
locate the transition as the state point where simple uniaxial order (solely described by a nonzero value of $S$) becomes unstable for a small perturbation by a finite value of the full biaxiality order parameter $F$, while setting $P=U=0$.
More details are given in Sec.~\ref{sec_perfectUAorder}
and the results of this method are included in our phase diagrams of hard isosceles spherotriangles in Fig.~\ref{fig_PD}, where we argue that it is quite reliable, since the uniaxial--uniaxial--biaxial triple points are consistently approached from both sides, where uniaxial order is assumed with respect to different main axes.
We thus expect that the assumption that one particle axis is perfectly aligned yields only reliable results  at unrealistically high packing fractions.

To compare our two approaches, we present  in Fig.~\ref{fig_UB} three phase diagrams of hard isosceles spherotriangles, each obtained upon imposing a different type of uniaxial order, as illustrated in Fig.~\ref{fig_coordinates}a, but here over the full range $0\leq x\leq 1$ of shape ratios.
Compared to the perturbative approach, the perfect uniaxial approximation systematically overestimates the packing fraction at the uniaxial--biaxial transition for the prolate uniaxial phases N$^\text{ph}$ and N$^\text{pb}$, while it is underestimated in the oblate case.
Therefore, when only considering the most stable ordered state (where the transition is predicted at the lowest density among the results for the three different uniaxial orientations), the perfect uniaxial approximation does not result in a consistent location of the triple points, in contrast to the perturbation result shown in Fig.~\ref{fig_PD}.

However, one noteworthy advantage of the perfect uniaxial approximation, as opposed to the perturbation approach, is its ability to produce closed analytic expressions for the transition densities (although these are, in general, too long to be stated here).
Only in the limits $x\rightarrow0$ and $x\rightarrow1$ of hard spherocylinders, the results can be presented in a compact form (recall the discussion in appendix~\ref{app_fullWDhsc}).
In the limit where the spherocylinder is perfectly aligned along its long axis, no biaxial order is possible as $F$ drops out of the functional, compare Eq.~\eqref{eq_wdscylinder}.
Hence, the results of our two methods  in Fig.~\ref{fig_UB} become comparable when this limit is approached.
In the (unphysical) opposite limit of the rod axis being perpendicular to its perfect orientation, $F$ remains in the functional, compare Eq.~\eqref{eq_wdscylinder2}, and the uniaxial--biaxial transition can be formally located at
\begin{equation}
    \eta_\text{IU}=\frac{3 \sqrt{1536 l^2+225 l^4+ 2304 l^3}-45 l^2-192l-128}{18 l^2-192l-128}\,,
\end{equation}
which is given as an explicit function of the aspect ratio~$l$.
 For $l=5$, we get $\eta_\text{IU}=0.255$, which lies below the (unstable) isotropic--uniaxial transition and is thus not shown in Fig. \ref{fig_UB}.

\end{document}